
\input harvmac
\newif\ifdraft

\noblackbox
\catcode`\@=11
\newif\iffrontpage
%
\ifx\answ\bigans
\def\titleft{\titsm}
\magnification=1200\baselineskip=15pt plus 2pt minus 1pt
%
\advance\hoffset by-0.075truein
\hsize=6.15truein\vsize=600.truept\hsbody=\hsize\hstitle=\hsize
\else\let\lr=L
\def\titleft{\titla}
\magnification=1000\baselineskip=14pt plus 2pt minus 1pt
%
\vsize=6.5truein
\hstitle=8truein\hsbody=4.75truein
\fullhsize=10truein\hsize=\hsbody
\fi
\parskip=4pt plus 10pt minus 4pt

\font\titla=cmr10 scaled\magstep3
\font\tenmss=cmss10
\font\absmss=cmss10 scaled\magstep1
\newfam\mssfam
\font\footrm=cmr8  \font\footrms=cmr5
\font\footrmss=cmr5   \font\footi=cmmi8
\font\footis=cmmi5   \font\footiss=cmmi5
\font\footsy=cmsy8   \font\footsys=cmsy5
\font\footsyss=cmsy5   \font\footbf=cmbx8
\font\footmss=cmss8
\def\footfont{\def\rm{\fam0\footrm}
\textfont0=\footrm \scriptfont0=\footrms
\scriptscriptfont0=\footrmss
\textfont1=\footi \scriptfont1=\footis
\scriptscriptfont1=\footiss
\textfont2=\footsy \scriptfont2=\footsys
\scriptscriptfont2=\footsyss
\textfont\itfam=\footi \def\it{\fam\itfam\footi}
\textfont\mssfam=\footmss \def\mss{\fam\mssfam\footmss}
\textfont\bffam=\footbf \def\bf{\fam\bffam\footbf} \rm}
\def\tenpoint{\def\rm{\fam0\tenrm}
\textfont0=\tenrm \scriptfont0=\sevenrm
\scriptscriptfont0=\fiverm
\textfont1=\teni  \scriptfont1=\seveni
\scriptscriptfont1=\fivei
\textfont2=\tensy \scriptfont2=\sevensy
\scriptscriptfont2=\fivesy
\textfont\itfam=\tenit \def\it{\fam\itfam\tenit}
\textfont\mssfam=\tenmss \def\mss{\fam\mssfam\tenmss}
\textfont\bffam=\tenbf \def\bf{\fam\bffam\tenbf} \rm}
\ifx\answ\bigans\def\abstractfont{\tenpoint}\else
\def\abstractfont{\def\rm{\fam0\absrm}
\textfont0=\absrm \scriptfont0=\absrms
\scriptscriptfont0=\absrmss
\textfont1=\absi \scriptfont1=\absis
\scriptscriptfont1=\absiss
\textfont2=\abssy \scriptfont2=\abssys
\scriptscriptfont2=\abssyss
\textfont\itfam=\bigit \def\it{\fam\itfam\bigit}
\textfont\mssfam=\absmss \def\mss{\fam\mssfam\absmss}
\textfont\bffam=\absbf \def\bf{\fam\bffam\absbf}\rm}\fi
%
\def\f@@t{\baselineskip10pt\lineskip0pt\lineskiplimit0pt
\bgroup\aftergroup\@foot\let\next}
\setbox\strutbox=\hbox{\vrule height 8.pt depth 3.5pt width\z@}
\def\vfootnote#1{\insert\footins\bgroup
\baselineskip10pt\footfont
\interlinepenalty=\interfootnotelinepenalty
\floatingpenalty=20000
\splittopskip=\ht\strutbox \boxmaxdepth=\dp\strutbox
\leftskip=24pt \rightskip=\z@skip
\parindent=12pt \parfillskip=0pt plus 1fil
\spaceskip=\z@skip \xspaceskip=\z@skip
\Textindent{$#1$}\footstrut\futurelet\next\fo@t}
\def\Textindent#1{\noindent\llap{#1\enspace}\ignorespaces}
\def\footnote#1{\attach{#1}\vfootnote{#1}}%

\def\foot{\attach\footsymbolgen\vfootnote{\footsymbol}}
\let\footsymbol=\star
\newcount\lastf@@t           \lastf@@t=-1
\newcount\footsymbolcount    \footsymbolcount=0
\def\footsymbolgen{\relax\footsym
\global\lastf@@t=\pageno\footsymbol}
\def\footsym{\ifnum\footsymbolcount<0
\global\footsymbolcount=0\fi
{\iffrontpage \else \advance\lastf@@t by 1 \fi
\ifnum\lastf@@t<\pageno \global\footsymbolcount=0
\else \global\advance\footsymbolcount by 1 \fi }
\ifcase\footsymbolcount \fd@f\star\or
\fd@f\dagger\or \fd@f\ast\or
\fd@f\ddagger\or \fd@f\natural\or
\fd@f\diamond\or \fd@f\bullet\or
\fd@f\nabla\else \fd@f\dagger
\global\footsymbolcount=0 \fi }
\def\fd@f#1{\xdef\footsymbol{#1}}
\def\space@ver#1{\let\@sf=\empty \ifmmode #1\else \ifhmode
\edef\@sf{\spacefactor=\the\spacefactor}
\unskip${}#1$\relax\fi\fi}
\def\attach#1{\space@ver{\strut^{\mkern 2mu #1} }\@sf\ }
%
\newif\ifnref
\def\rrr#1#2{\relax\ifnref\nref#1{#2}\else\ref#1{#2}\fi}
\def\ldf#1#2{\begingroup\obeylines
\gdef#1{\rrr{#1}{#2}}\endgroup\unskip}
\def\nrf#1{\nreftrue{#1}\nreffalse}
\def\doubref#1#2{\refs{{#1},{#2}}}

\nreffalse
\def\refout{\listrefs}
%
\def\eqn#1{\xdef #1{(\secsym\the\meqno)}
\writedef{#1\leftbracket#1}%
\global\advance\meqno by1\eqno#1\eqlabeL#1}
\def\eqnalign#1{\xdef #1{(\secsym\the\meqno)}
\writedef{#1\leftbracket#1}%
\global\advance\meqno by1#1\eqlabeL{#1}}
%
\def\chap#1{\newsec{#1}}
\def\chapter#1{\chap{#1}}
\def\sect#1{\subsec{{ #1}}}
\def\section#1{\sect{#1}}
\def\\{\ifnum\lastpenalty=-10000\relax
\else\hfil\penalty-10000\fi\ignorespaces}
\def\note#1{\leavevmode%
\edef\@@marginsf{\spacefactor=\the\spacefactor\relax}%
\ifdraft\strut\vadjust{%
\hbox to0pt{\hskip\hsize%
\ifx\answ\bigans\hskip.1in\else\hskip-.1in\fi%
\vbox to0pt{\vskip-\dp
\strutbox\sevenbf\baselineskip=8pt plus 1pt minus 1pt%
\ifx\answ\bigans\hsize=.7in\else\hsize=.35in\fi%
\tolerance=5000 \hbadness=5000%
\leftskip=0pt \rightskip=0pt \everypar={}%
\raggedright\parskip=0pt \parindent=0pt%
\vskip-\ht\strutbox\noindent\strut#1\par%
\vss}\hss}}\fi\@@marginsf\kern-.01cm}
\def\titlepage{%
\frontpagetrue\nopagenumbers\abstractfont%
\hsize=\hstitle\rightline{\vbox{\baselineskip=10pt%
{\abstractfont\pubnum}}}\pageno=0}
\frontpagefalse
\def\pubnum{}
\def\pdate{\number\month/\number\yearltd}
\def\makefootline{\iffrontpage\vskip .27truein
\line{\the\footline}
\vskip -.1truein\leftline{\vbox{\baselineskip=10pt%
{\abstractfont\pdate}}}
\else\vskip.5cm\line{\hss \tenrm $-$ \folio\ $-$ \hss}\fi}
\def\title#1{\vskip .7truecm\titlestyle{\titleft #1}}
\def\titlestyle#1{\par\begingroup \interlinepenalty=9999
\leftskip=0.02\hsize plus 0.23\hsize minus 0.02\hsize
\rightskip=\leftskip \parfillskip=0pt
\hyphenpenalty=9000 \exhyphenpenalty=9000
\tolerance=9999 \pretolerance=9000
\spaceskip=0.333em \xspaceskip=0.5em
\noindent #1\par\endgroup }
\def\autskip{\ifx\answ\bigans\vskip.5truecm\else\vskip.1cm\fi}
\def\author#1{\vskip .7in \centerline{#1}}

\def\address#1{\ifx\answ\bigans\vskip.2truecm
\else\vskip.1cm\fi{\it \centerline{#1}}}
\def\abstract#1{
\vskip .5in\vfil\centerline
{\bf Abstract}\penalty1000
{{\smallskip\ifx\answ\bigans\leftskip 2pc \rightskip 2pc
\else\leftskip 5pc \rightskip 5pc\fi
\noindent\abstractfont \baselineskip=12pt
{#1} \smallskip}}
\penalty-1000}
\def\endpage{\tenpoint\supereject\global\hsize=\hsbody%
\frontpagefalse\footline={\hss\tenrm\folio\hss}}
\def\ack{\goodbreak\vskip2.cm\centerline{{\bf Acknowledgements}}}

\def\bfone{\relax{\rm 1\kern-.35em 1}}
\def\inbar{\vrule height1.5ex width.4pt depth0pt}
\def\IC{\relax\,\hbox{$\inbar\kern-.3em{\mss C}$}}
\def\ID{\relax{\rm I\kern-.18em D}}
\def\IF{\relax{\rm I\kern-.18em F}}
\def\IH{\relax{\rm I\kern-.18em H}}
\def\II{\relax{\rm I\kern-.17em I}}
\def\IN{\relax{\rm I\kern-.18em N}}
\def\IP{\relax{\rm I\kern-.18em P}}
\def\IQ{\relax\,\hbox{$\inbar\kern-.3em{\rm Q}$}}
\def\IR{\relax{\rm I\kern-.18em R}}
\font\cmss=cmss10 \font\cmsss=cmss10 at 7pt
\def\ZZ{\relax\ifmmode\mathchoice
{\hbox{\cmss Z\kern-.4em Z}}{\hbox{\cmss Z\kern-.4em Z}}
{\lower.9pt\hbox{\cmsss Z\kern-.4em Z}}
{\lower1.2pt\hbox{\cmsss Z\kern-.4em Z}}\else{\cmss Z\kern-.4em Z}\fi}
\def\a{\alpha} \def\b{\beta} \def\d{\delta}
\def\e{\epsilon} \def\c{\gamma}
 \def\l{\lambda}
 \def\s{\sigma}
\def\cA{{\cal A}} 
 \def\cD{{\cal D}}
\def\cF{{\cal F}} 
 
 \def\cK{{\cal K}}
 \def\cM{{\cal M}}
 \def\cO{{\cal O}}
 \def\cQ{{\cal Q}}
\def\cR{{\cal R}} 
\def\nup#1({Nucl.\ Phys.\ $\us {B#1}$\ (}
\def\plt#1({Phys.\ Lett.\ $\us  {#1}$\ (}
\def\cmp#1({Comm.\ Math.\ Phys.\ $\us  {#1}$\ (}
\def\prp#1({Phys.\ Rep.\ $\us  {#1}$\ (}
\def\prl#1({Phys.\ Rev.\ Lett.\ $\us  {#1}$\ (}
\def\prv#1({Phys.\ Rev.\ $\us  {#1}$\ (}
\def\mpl#1({Mod.\ Phys.\ \Let.\ $\us  {#1}$\ (}
\def\tit#1|{{\it #1},\ }
%

%

\def\ni{\noindent}
\def\tilde{\widetilde}
\def\bar{\overline}
\def\us#1{\underline{#1}}
\let\shat=\hat
\def\hat{\widehat}

\def\Coe#1.#2.{{#1\over #2}}
\def\coeff#1#2{\relax{\textstyle {#1 \over #2}}\displaystyle}
\def\coe#1.#2.{\relax{\textstyle {#1 \over #2}}\displaystyle}
\def\half{{1 \over 2}}
\def\shalf{\relax{\textstyle {1 \over 2}}\displaystyle}

\def\to{\rightarrow}
\def\notin{\hbox{{$\in$}\kern-.51em\hbox{/}}}

\def\exx#1{e^{{\displaystyle #1}}}
\def\del{\partial}

\def\nex#1{$N\!=\!#1$}

\catcode`\@=12
%
\def\cS{{\cal K}}
\def\IE{\relax{{\rm I\kern-.18em E}}}
\def\cE{{\cal E}}
\def\rt{{\cR^{(3)}}}
\def\IGam{\relax{{\rm I}\kern-.18em \Gamma}}
\def\IGa{\IA}
\def\LG{Lan\-dau-Ginz\-burg\ }

\def\Rt{{\cal R}^{(3)}}
\def\wabc{W_{abc}}
\def\WABC{W_{\a\b\c}}
\def\W{{\cal W}}
\def\tft#1{\langle\langle\,#1\,\rangle\rangle}
\def\IA{\relax{\mss  A}}

\def\hata{{\shat\a}}
\def\hatb{{\shat\b}}
\def\hatA{{\shat A}}
\def\hatB{{\shat B}}

\def\bv{{\bf V}}
\def\spg{special geometry}
\def\sc{SCFT}
\def\leel{low energy effective Lagrangian}
\def\pf{Picard--Fuchs}
\def\el{effective Lagrangian}
\def\Fb{\overline{F}}

\def\Ub{\overline{U}}

\def\zb{\overline{z}}

\def\tb{\overline{t}}
\def\Xb{\overline{X}}
\def\Vb{\overline{V}}

\def\cy{Calabi--Yau}

\def\Kh{\hat{K}}
\def\Knh{{\cal K}}

\def\V{{V}}
\def\T{T}
\def\Gammah{\hat{\Gamma}}

\def\K{K\"ahler}

\def\w{w}
%
%
\ldf\FL{S.\ Ferrara and J. Louis, {\it Flat holomorphic
   connections and Picard-Fuchs identities from
   N=2 supergravity}, preprint CERN-TH.6334/91.}
\ldf\LSW{W.\ Lerche, D.\ Smit and N.\ Warner, {\it Differential equations for
   periods and flat coordinates in two-dimensional topological matter
theories},
 preprint LBL-31104,USC-91/022,CALT-68-1738.}
\ldf\Kos{B.\ Kostant, Am.\ J.\ Math.\ 81 (1959) 973.}
\ldf\Ferr  {A.\ Cadavid and S.\ Ferrara, \plt B267 (1991) 193.}
\ldf\DIZ  {P.\ Di Francesco, C.\ Itzykson and J.-B.\ Zuber, \cmp140 (1991)
543.}
\ldf\Forsyth  {A. Forsyth, {\it Theory of Differential Equations}, Vol. 4,
Dover Publications, New-York (1959).}
\ldf\Vafa{C.\ Vafa, Mod.\ Phys.\ Lett. A6 (1991) 337.}
\ldf\DVV {R.\ Dijkgraaf, E. Verlinde and H. Verlinde,
\nup352(1991) 59.}
\ldf\CandB  {P.\ Candelas, X.C.\ de la Ossa, P.S.\ Green and
L.\ Parkes, \plt 258B (1991) 118; \nup359 (1991) 21.}
\ldf\Strom  {A.\ Strominger, \cmp133 (1990) 163.}
\ldf\BVar  {B. Blok and  A.\ Varchenko, \tit Topological
 conformal field theories and the flat coordinates| preprint
 IASSNS-HEP-91/5.}
\ldf\CeVa  {S.\ Cecotti and C.\ Vafa, \nup367 (1991) 359.}
\ldf\LVW{W.\ Lerche, C.\ Vafa and N.P.\ Warner, \nup324(1989) 427.}
\ldf\CandA{P.\ Candelas, \nup 298 (1988) 458.}
\ldf\FLT{S.\ Ferrara, D.\ L\"ust and S.\ Theisen,  \plt 242B (1990) 39.}
\ldf\cubicF{E.\ Cremmer, C.\ Kounnas, A.\ Van Proeyen, J.\ Derendinger,
S.\ Ferrara, B.\ de Wit and L.\ Girardello, \nup250 (1985) 385 \semi
   other cubic F-theories, de Wit xxxxxx?????}
\ldf\Dublin{J.\ Balog, L.\ Feher, L.\ O'Raifeartaigh, P.\ Forga\'cs
and A.\ Wipf, \plt244B (1990) 435; Ann.\ Phys.\ 203 (1990) 194.}
\ldf\Cec{S.\ Cecotti,
 \nup 355 (1991) 755, Int.\ J.\ Mod.\ Phys.\ A6 (1991) 1749.}
\ldf\GiSm{A.\ Giveon and  D.-J.\ Smit, Mod. Phys. Lett. A {\bf 6}
No. 24 (1991) 2211.}
\ldf\Katz{N.\ Katz, Publ.\ Math.\ I.H.E.S.\ 35 (1968) 71.}
\ldf\DM{D.~Morrison,
{\it \pf\ equations and mirror maps for hypersurfaces},
 Duke preprint DUK-M-91-14 (1991).}
\ldf\Bott{M.F.\ Atiyah, R.\ Bott anf L.\ G\"aarding, Acta
Mathematica 131 (1973) 145.}
\ldf\Griff{See eg., P.\ Griffiths, Ann.\ Math.\ 90 (1969) 460.}
\ldf\Fre{P.\ Fr\'e and P.\ Soriani, {\it Symplectic embeddings, K\"ahler
geometry and automorphic functions: The Case of $SK(n+1) =
   SU(1,1) / U(1) \times SO(2,n) / SO(2) \times SO(n)$}, preprint SISSA
 90/91/EP; S.\ Ferrara, P.\  Fr\'e and P.\ Soriani, {\it On the moduli space
of the $T^6/Z_3$ orbifold and its modular group}, preprint CERN-TH.6364/92,
SISSA 5/92/EP.}
\ldf\GVW{B. Greene, C.\ Vafa and N.P.\ Warner, \nup324 (1989) 371. }
\ldf\WL{W.\ Lerche, \nup 238 (1984) 582; W.\ Lerche and W.\ Buchm\"uller,
Ann.\ Phys.\ 175 (1987) 159.}
\ldf\DKL{L.J.\ Dixon, V.S.\ Kaplunovsky and J.\ Louis,
  \nup329 (1990) 27.}
\ldf\dfnt{B.\ de Wit and A.\ Van Proeyen, \nup245 (1984) 89;
B.\ de Wit, P.\ Lauwers and A.\ Van Proeyen, \nup255 (1985) 569;
E.\ Cremmer, C.\ Kounnas, A. \ Van Proeyen, J.P.\ Derendinger,
S.\ Ferrara, B.\ de Wit and L.\ Girardello,  \nup250 (1985) 385.
}
\ldf\CO{P.\ Candelas and X.C.\ de la Ossa, \nup355 (1991) 455.}
\ldf\pftop{E.\ Verlinde and N.P.\ Warner, \plt 269B (1991) 96;
Z. Maassarani, \plt273B (1991) 457; A.~Klemm, M.~G.~Schmidt and S.~Theisen,\
{\it Correlation functions
for topological Landau-Ginzburg models with $c\leq3$}, Karlsruhe preprint
KA-THEP-91-09 (1991).}
\ldf\tntwo{E.\ Witten, \cmp118 (1988) 411, \nup340 (1990) 281; T.\ Eguchi and
 S.K.\ Yang, Mod.\ Phys.\ Lett.\ A5 (1990) 1693.}
\ldf\ntwo{T.\ Banks, L.\ Dixon, D.\ Friedan and S.\ Shenker, \nup299 (1988)
 613.}
\ldf\AMW{P.\ Aspinwall and D.\ Morrison, {\it Topological field theory and
 rational curves}, preprint DUK-M-91-12; E.\ Witten, {\it Mirror manifolds
and topological field theory}, preprint IASSNS-HEP-91/83.}
\ldf\Ferrara{S.\ Ferrara and A.\ Strominger, {\it N=2 spacetime
supersymmtry and Calabi-Yau moduli space}, preprint CERN-TH.5291/89;
S.\ Cecotti, \cmp131(1990) 517;
A.\ Cadavid, M.\ Bodner and S.\ Ferrara, \plt B247 (1991) 25.}
\ldf\fetal{L.~Castellani, R.~D'Auria and S.~Ferrara,
    \plt B241 (1990) 57; Class.\ Quant.\ Grav.\ 1 (1990) 317; R.~D'Auria,
S.~Ferrara and P.~Fr\'e, \nup359 (1991) 705.}
\ldf\mirror{L.\ Dixon and D.\ Gepner, unpublished; W.\ Lerche, C.\ Vafa and
N.P.\ Warner, \nup324(1989) 427; B.\ Greene and M.\ Plesser, \nup 338 (1990)
15; P.\ Candelas, M.\ Lynker and R.\ Schimmrigk, \nup 341 (1990) 383.}
\ldf\ns{N.~Seiberg, \nup303 (88) 206.}
\ldf\cfg{
S.\ Cecotti, S.\ Ferrara and L.\ Girardello,
Int.\ Mod.\ J.\ Phys.\ A4 (1989) 2475; \plt B213 (1988) 443.}
\ldf\cubicF{E.~Cremmer and A.~Van Proeyen, Class. Quant. Grav.
{\bf 2} (1985) 445;
S.\ Cecotti, \cmp124(1989) 23;
B.~de Wit and A.~Van Proeyen, {\it Special geometry, cubic polynomials
and homogeneous quaternionic spaces},
CERN--preprint  TH.6302/91.}
\ldf\DS{V.G.\ Drinfel'd and V.~V.~Sokolov,
 Jour.~Sov.~Math. {\bf 30} (1985) 1975.}
\ldf\Font{A.\ Font, {\it Periods and Duality Symmetries in \cy\
Compactifications}, preprint UCVFC/DF-1-92.}
\def\Symp{(A.25)}
\def\aff#1#2{\centerline{$^{#1}${\it #2}}}
\def\pubnum{
\hbox{CERN-TH.6441/92}
}
\def\pubnum{
\hbox{CERN-TH.6441/92}
\hbox{UCLA/92/TEP/8}
\hbox{CALT-68-1776}
\hbox{POLFIS-TH. 08/92}
}
\def\pdate{
\hbox{hepth@xxx/9204035}
\hbox{CERN-TH.6441/92}
\hbox{March 1992}
}
\titlepage
\title
 {Picard--Fuchs Equations and Special Geometry\foot{Supported in part by DOE
grants DE-AC0381-ER50050, DOE-AT03-88ER40384,Task E, and by M.U.R.S.T.\ .}}
\vskip-.8cm
\author{
A.\ Ceresole$^{1,5}$,
R.\ D'Auria$^2$,
S.\ Ferrara$^{3,4}$,
W.\ Lerche$^3$ and J.\ Louis$^3$}
\vskip2.truecm
\aff1{California Institute of Technology, Pasadena, CA 91125, USA}
\aff2{Dipartimento di Fisica, Politecnico di Torino
and INFN, Sezione di Torino, Italy}
\aff3{CERN, 1211 Geneva 23, Switzerland}
\aff4{Department of Physics, UCLA,
Los Angeles, USA.}
\aff5{Dipartimento di Fisica Teorica, Universit\`a di Torino and INFN,}
\aff{}{Sezione di Torino, Italy.}

\vskip-.8 cm
\abstract
{\ni We investigate the system of holomorphic differential identities implied
by special K\"ahlerian geometry of four--dimensional $N=2$ supergravity. For
superstring compactifications on \cy\ threefolds these identities are
equivalent to the Picard--Fuchs equations of algebraic geometry that are obeyed
by the periods of the holomorphic three--form. For one variable they reduce to
linear fourth--order equations which are characterized by classical
$W$--generators; we find that the instanton corrections to the Yukawa couplings
are directly related to the non--vanishing of $w_4$. We also show that the
symplectic structure of special geometry can be related to the fact that the
Yukawa couplings can be written as triple derivatives of some holomorphic
function $F$. Moreover, we give the precise relationship of the Yukawa
couplings of special geometry with three--point functions in topological field
theory.}

\endpage

\chap{Introduction and Summary}

It is well known that
four--dimensional heterotic string vacua which are $N=1$ space--time
supersymmetric have necessarily an $N=2,c=9$ \sc\ in the left--moving sector
\ntwo. Furthermore, the couplings of the corresponding \leel\ are directly
related to correlation functions in the $N=2$ \sc. Unfortunately, for generic
string vacua we are currently not able to calculate the relevant correlation
functions. However, recently for a particular family of non--trivial string
vacua (a compactification on a specific \cy\ threefold), some of the couplings
in the \leel\ have been computed exactly (at the string tree level, but to all
orders in the $\sigma$--model coupling) by using techniques of
algebraic geometry and without ever relying on the underlying \sc\ \CandB. It
is clearly important to understand in some detail the general structure behind
this specific example. Indeed, it was shown in ref.~\CandB\ that the couplings
could be obtained from the solution of a certain fourth--order linear
holomorphic differential equation. It was realized that this
differential equation is a particular case of ``\pf\ equations'' obeyed by the
periods of the holomorphic three--form $\Omega$ that exists on any \cy\
threefold
\nrf{\Ferr\LSW\DM} \refs{\Ferr{--}\DM}.
(\pf\ equations can be derived for general
``\cy'' $d$--folds \doubref\LSW\Ferr, but we consider only $d=3$ in the
following.)

In a parallel development, twisted $N=2$ \sc s were investigated \tntwo. It was
shown that the twisting essentially leads to a projection onto the massless
subsector of the original \sc\ and that certain correlation functions are
topological and (almost) identical to their untwisted analogues. In these
theories one can derive from consistency considerations \doubref\DVV\pftop\
differential equations that are equivalent to the \pf\ equations. We should
note that not only the \pf\ equations arise from these topological
considerations but there are further properties of the \leel\ encoded in
some appropriate topological field theory (TFT) \doubref\AMW\CeVa.

A further step in uncovering the general structure behind the differential
equation was undertaken in ref.~\FL. It was realized that the \pf\ equations
for a \cy\ threefold are just another way of expressing a geometrical structure
called ``special geometry'' \nrf{\dfnt\Ferrara\Strom\CO\fetal}
\refs{\dfnt{--}\fetal}. It first arose in the study of coupling vector
multiplets
to $N=2$ supergravity in four dimensions \dfnt. In string theory, special
geometry is related to the subclass of string vacua with $(2,2)$ worldsheet
supersymmetry\nrf{\ns\cfg\DKL} \refs{\ns{--}\DKL}. The additional
right--moving world--sheet supersymmetry implies further constraints on the
couplings of the \el. In particular, in the moduli
sector of the theory these constraints can be expressed by the equations
\refs{\dfnt{,}\Strom{--}\fetal{,}\DKL}
$$
\eqalign{
R_{\alpha \bar \beta \gamma}^{ \delta} = &
g_{\alpha \bar \beta} \delta_{\gamma}^{\delta} +
g_{\gamma \bar \beta} \delta_\alpha^\delta     -
C_{\alpha\gamma\epsilon} g^{\epsilon \bar \epsilon}
C_{\bar \beta \bar \delta \bar \epsilon}  g^{\delta\bar \delta}
\, ,  \cr
C_{\alpha\beta\gamma} =& \
e^K W_{\alpha\beta\gamma}(z) \, .}
\eqn\cwrelation
$$
Here, $g_{\alpha \bar \beta} (z,\bar z) = \del_\alpha {\del}_{\bar \beta}
K(z,\bar z)$ is the \K\ metric on the moduli space and the (completely
symmetric) Yukawa couplings $W_{\alpha\beta\gamma}$ are holomorphic functions
of the moduli $z^\alpha$ ($\a=1,\dots,n$, where $n$ is the dimension of the
moduli space). A complex \K\ manifold whose metric obeys \cwrelation\ is called
a special \K\ manifold. (Further relevant formulas of special geometry
are collected in Appendix A.)

For given $W_{\a\b\c}$, eq.~\cwrelation\ can be viewed as a covariant and
non-holomorphic differential equation for the \K\ potential. Its general
solution can be expressed \doubref\fetal\Strom\
in terms of $n+1$ holomorphic sections $X^A(z), A =
0,1,\ldots,n$ which obey ${\del}_{\bar \alpha} X^A = 0$:
$$
K\ =\ -\ln i (X^A \Fb_{A} - \Xb^A F{_A} )\, ,  \eqn\ksolK
$$ where $$
W_{\alpha\beta\gamma}\ =\  \del_\alpha X^A \del_\beta X^B
\del_\gamma X^C F_{ABC}\, ,\eqn\ksolW
$$
and
$
F_{A}(X) = {\del F(X)\over \del X^A  }
$, {\it etc},
and $F(X)$ is a homogeneous function of $X$ of degree two. We see that all
information about $K$ and $W_{\alpha\beta\gamma}$ is encoded in the
holomorphic objects $X^A(z)$ and $F_A(z)$ and their complex conjugates.

In order to make contact with the differential equation of ref.~\CandB, one
observes \refs{\fetal{,}\Strom{,}\CO}
that eq.~\cwrelation\ is entirely equivalent to the following system of
non--holomorphic first--order equations
$$
\eqalign{
D_\alpha V &= U_\alpha \cr
D_\alpha U_\beta &= - i C_{\alpha\beta\gamma} g^{\gamma \bar \gamma}
\bar U_{\bar \gamma} \cr
D_\alpha \bar U_{\bar \beta} &= g_{\alpha \bar \beta} \Vb \cr
D_\alpha \Vb &= 0 \, ,                               }
\eqn\deqset
$$
where $V(z)= \left(X^A(z),  F_A (z)\right)$ and $D_\alpha$
 is the \K\ and
reparametrization covariant derivative.
By successively
inserting these equation into each other one can represent
this system by
$$
D_\a D_\b ({C^{{-1}{\shat\c}}})^{\rho\sigma}D_{\shat\c}D_\s V=0
\eqn\ins
$$
(assuming for the moment that the matrix $(C_\a)_{\b\c}$ is invertible). Here,
$\hat\c$ is a priori not summed over (in contrast to $\s$), but it is easy to
show that the equation remains valid even if also $\hat\c$ is summed over. In
ref.~\FL\ it was shown that eq.~\ins\ in one complex dimension ($n$=1) is
actually holomorphic, although its building blocks are not.
It is the covariant
version of the (holomorphic) equation of ref.~\CandB, and thus is the analogue
of the \pf\ equations in special geometry. Its solution determines $X^A$ and
$F_A$ and thus via eqs.~\ksolW,\ksolK\ also $W_{\alpha\beta\gamma}$ and $K$.

The existence of the covariant holomorphic differential equation \ins\ is
intimately connected to the fact that the Christoffel as well as the
\K\ connection naturally split into the sum of two terms \FL. One of them is
non--holomorphic and transforms as a tensor whereas the other term is
holomorphic and transforms like a connection. Furthermore, the holomorphic
pieces of these connections are flat and vanish in ``special coordinates''
$$
t^a(z) = {X^a(z) \over X^0(z)} \ \qquad\ \ \  (a=1,\dots,n)\ .
\eqn\specoo
$$
A similar situation holds in topological \LG models where the flat
connection can be identified with the Gauss--Manin connection
\nrf{\BVar} \refs{\BVar,\CeVa,\LSW}.

 Ref.~\FL\ studied equation~\ins\ in detail only in one complex dimension; in
this paper we generalize the analysis to arbitrary dimensions $n$. However, we
start in section~2 by adding a few observations to the results of ref.~\FL. In
particular we show that eq.~\ins\ in one dimension is not the most general
linear fourth--order differential equation but rather is characterized by the
vanishing of one of the ``invariants'', $w_3=0$. The other invariant $w_4$
measures the deviation from covariantly constant\foot{Covariantly constant with
respect to the holomorphic connections.} Yukawa couplings, or in
other words, the deviation from the large radius (classical) limit of the \cy\
moduli space.

Every $N$--th order differential equation is equivalent to a first--order
matrix
equation of the form $(\del - \IA)\bv=0$, where the first row of $\bv$ is the
solution vector $V$ in \ins. In section 2.2 we show that $w_3=0$ translates
into the statement that the gauge potential $\IA$ of this matrix equation takes
values in $sp(4)$. It is this fact which nicely generalizes to the case of an
$n$--dimensional moduli space.

In section~3 we derive the \pf\ equations of special geometry for an arbitrary
number of moduli. They are a direct consequence of \ksolW, and are most easily
displayed as $n$ coupled first--order holomorphic matrix
equations:
$$(\del_\alpha - \IA_\alpha)\bv\ =\ 0\ .
\eqn\meq
$$
Here, $\IA$ takes values in $sp(2n+2)$. This is the analogue of the vanishing
of
$w_3$ in one dimension. $\IA_\alpha$ is a sum of a matrix $\IGam_\alpha$ which
contains the flat connections plus the structure constants ${\IC}_\alpha$ of a
$2n+2$ dimensional chiral ring $\rt$. The structure constants contain the
Yukawa couplings $W_{\alpha\beta\gamma}$ and furthermore satisfy
$[\IC_\alpha,\IC_\beta]=0,\, \IC^4 =0$. Because the connection is symplectic,
$\bv$ can always be taken as an element of $Sp(2n+2)$. This means that the
well--known symplectic structure of special geometry can ultimately be traced
back to the identity \ksolW.

In section 3.2 we display the relationship between equations \meq\ and \deqset.
Eq.~\deqset\ can also be written as a first--order matrix equation
$(\del_\alpha - \cA_\alpha){\bf U}=0$, albeit with a non-holomorphic connection
$\cA$. Strominger observed \Strom\ that the system \cwrelation\ is just the
flatness condition for $\cA$. We will show that eq.~\meq\ corresponds to a
gauge where $\bar{\cA} = 0, \cA = \IA, \bar\del\IA=0$.

In section 3.3 we consider
particular cases where ${\IC}$ is degenerate, which corresponds to
decoupled chiral rings. Here the $F$--function is a direct sum whereas the
metric of the moduli space is a complicated function and by no means the metric
on a product space. This clearly shows that the fundamental object in special
geometry is indeed the holomorphic function $F$ and not the non-holomorphic
metric on moduli space.

As we indicated above, the motivation for the present work was to analyse the
holomorphic differential equations of special geometry. So far, the discussion
has been completely general and applies also to geometries that do not have an
interpretation in terms of Calabi-Yau manifolds or TFT's. In section 4 we
relate eqs.~\deqset--\meq\ to \cy\ moduli spaces, where $V, U_\alpha,
\Ub_{\bar{\alpha}}, \Vb$ correspond to basis elements of the third cohomology,
$H^3= H^{(3,0)} \oplus H^{(2,1)} \oplus H^{(1,2)} \oplus H^{(0,3)}$
\doubref\Strom\CO. We then continue to relate the formulas used in topological
\LG\ theory to the structures uncovered in \spg\ in section~3.
This is useful in order to explicitly compute the differential equations.
In particular we verify that the computation of Yukawa couplings
$\WABC$ via \pf\
equations in \spg\ is, as expected, identical to the computation of certain
three--point correlators in topological \LG theory.

We should stress that in the context of special geometry the deformations of
the \K\ class and the deformations of the complex structure appear on an equal
footing. This is due to the fact that eq.~\cwrelation\ holds for both types of
moduli \refs{\DKL,\ns} which is another manifestation of ``mirror symmetry''
\refs{\mirror{,}\CandB{,}\AMW}. In practical applications it is often possible
to compute the \pf\ equations for one type of moduli only. In order to find the
\pf\ equation for the other class of moduli, one needs to make use of the
mirror symmetry.

By using ``topological anti--topological fusion'' the geometrical structure
implied by eq.~\cwrelation\ was extended \CeVa\ to include relevant (massive)
perturbations in addition to the marginal (massless) moduli considered here.
One of the main objectives of \CeVa\ was to construct non--holomorphic
quantities (like the metric) from TFT. On the other hand, the emphasis of the
present paper is on the structure of the holomorphic \pf\ equations, in
relation to \spg.

Moreover, in Appendix C we discuss the differential equations for cubic
$F$--functions.

\chap{Differential equations for one variable}
\sect{Linear differential equations and $W$--generators}

In ref.~\CandB\ it was shown that the periods of a one--dimensional moduli
space of a particular \cy--threefold (a quintic in $CP_4$) are determined by a
fourth--order linear differential equation. The corresponding differential
equation in special geometry ---the one--dimensional version of eq.~\ins--- was
derived in ref.~\FL. In this section we add some observations concerning this
one--dimensional case. This will prove advantageous for the study of the
general situation.

Thus, let us first briefly review some facts about linear fourth--order
differential equations \Forsyth. Their general form  reads
$$
 \sum_{n=0}^4 a_n(z) \, \del_z^n V\ =\ 0  \ ,
\eqn\oddeq
$$
where the $a_n$ obey well--defined transformation laws in order to render
eq.~\oddeq\ covariant under coordinate changes $z\to\tilde z(z)$,
$\del\to\xi^{-1}\del$, $\xi\equiv\del \tilde z/\del z$. Not all of the $a_n$
are relevant. First, one can scale out $a_4$, and furthermore drop the
coefficient proportional to $a_3$ by means of the redefinition
$
V\to \V e^{ - 1/4 \int {a_3(u)\over a_4(u)} du}
$.
This puts the differential equation into the form
$$
\cD V \equiv
(\del^4 + c_2 \del^2 + c_1 \del + c_0)V = 0  \ ,
\eqn\oddeqp
$$
where the new coefficients $c_n$ are combinations of the $a_n$ and their
derivatives. In this basis $V$ transforms as a $-3/2$ differential, but the
transformation properties of the $c_n$ are not very illuminating. However, one
can find combinations of the $c_n$'s and their derivatives which
transform like tensors:
 $$
\eqalign{
\w_2 = &\  c_2   \cr
\w_3 = &\  c_1 -  c_2' \cr
\w_4 = &\   c_0 - \coeff12  c_1'
+ \coeff{1}{5}   c_2''
- \coeff{9}{100}   c_2^2 \ .}
\eqn\wcoeff
$$
A straightforward computation shows
$$
\eqalign{
\tilde\w_2 =&\ \xi^{-2} [ \w_2 - 5\{\tilde z ;z\} ] \cr
\tilde\w_3 = &\ \xi^{-3} \w_3 \cr
\tilde\w_4 = &\ \xi^{-4} \w_4 \, ,}
\eqn\wtrans
$$
where $\{\tilde z ; z \}= ( { \del^2 \xi \over \xi} - {3\over2}({\del
\xi\over\xi})^2)$ is the Schwarzian derivative. Actually $w_2,w_3,w_4$ form a
classical $W_4$--algebra (see, for instance, \DIZ), a fact that we will not
make use of in this paper. Using \oddeqp\ and \wcoeff\ one finds
$$
\cD\,V =
\big[\del^4+ w_2 \del^2+(w_3+w_2')\del+
\coeff3{10}w_2''+\coeff9{100}{w_2}^2+\shalf w_3'
+ w_4\big]\,V   \, .
\eqn\wdeq
$$
The advantage of rewriting a differential equation in terms of $W$-generators
is that this is a convenient way to display the particular properties of the
equation in a reparametrization-covariant way. {}From eq.~\wtrans\ we learn
that there is always a coordinate system in which $w_2=0$ holds. On the other
hand, $w_3$ and $w_4$ do characterize the fourth--order differential operator
$\cD$ in any coordinate frame.

Let us return to special geometry:
in ref.~\FL\ it was shown that there is a holomorphic
fourth--order differential equation that expresses the constraint
of special geometry and thus is equivalent to eq.~\cwrelation.
This equation is the one-dimensional version of \ins\
and reads
$$
DD\,W^{-1}\,DDV\ =\ 0     \, ,\eqn\sgeq
$$
where $D$ is the K\"ahler-- and reparametrization covariant
derivative defined in (A.11), and $W$ is the
one-dimensional Yukawa coupling. In special coordinates \specoo,
this equation becomes very simple,
$$
\del^2\,W^{-1}\,\del^2\,V\ =\ 0\ .\eqn\simpdeq
$$
Equation \sgeq\ can be written in the form
\oddeq\ and one finds that the coefficients
are not arbitrary but are related as follows \FL:
$a_3 = 2 \del a_4, \,a_4 = W^{-1},
a_1 = \del a_2 - \coeff12 \del^2 a_3.$
The coefficients $a_2$ and $a_0$ are complicated functions of $W$ and the
connections. The above relations translate into the invariant
statement\foot{This was noted in \LSW\ for the special
 case of the quintic hypersurface in $CP^4$.}
$$
\w_3\ \equiv\ 0\ .
\eqn\wtzero
$$
Furthermore, the other $W$-generators are given (in special coordinates) by
$$
\eqalign{
w_2\ &=\ \coeff1{2W^2}\big({4 W W''-5 {{W'}^2}}\big)\cr
w_4\ &=\ \coeff1{100W^4}
  \big(175 {{W'}^4} - 280 W {{W'}^2} W'' + 49 {W^2} {{W''}^2} +
      70 {W^2} W' W'''\phantom{\big)}\cr
      &\qquad\qquad \phantom{\big(}- 10 {W^3} W''''\big)\ .
\cr}\eqn\wW
$$
Thus, all special geometries in one dimension lead to a fourth--order
linear differential equation that is characterized by  $w_3=0$.
This is in close relation to the fact that the solution vector $V$
does not consist of four completely independent elements, but rather
has a restricted structure. More precisely, by construction four linear
independent solutions are given by the components of the vector (cf., (A.22))
$$
V= \left(X^A(z),F_A(z)\right)\ , \qquad A=0,1 \, ,
\eqn\Vsol
$$
with
$$
F_A(z)\ = \ {\del\over \del X^A(z)} F(z)  ,\eqn\Fadef
$$
where $F$ is a homogeneous function of $X$ of degree 2.
The reverse statement is however not true:  $w_3=0$ does not imply
that the solution $V$ can always be written in the form
\Vsol. This is proved in Appendix B.

Note that the property \Fadef\ does not uniquely fix
$V$. It is known \refs{\dfnt{,}\CO{,}\fetal{,}\cfg}
that precisely for symplectic rotations of $V$,
$$
\big(\tilde X^A , \tilde F_A(\tilde X^A)\big)\ =\
\big(X^A, F_A(X^A)\big)\cdot M\ ,\qquad M\in Sp(4)\ ,
\eqn\spaction
$$
one has $\tilde F_A = (\del \tilde F/\del \tilde X^A) $ where $\tilde F$ is
again a homogeneous function of degree 2. Thus, the elements of $V$ are defined
only up to this kind of transformations. Of course, generic linear combinations
of the four solutions are still solutions of \sgeq, but for these the special
structure of the solutions (that reflects $w_3=0$) is not manifest.

Symplectic transformations belonging to $Sp(2n+2,\IR)$ have a particular
meaning in special geometry. They represent changes of special coordinate bases
and are exactly those transformations which leave $K$
form--invariant and consequently do not change any physical quantity. (This can
be easily seen from eq.~(A.24) which displays manifestly the symplectic
structure of $K$.) We will show in the following sections how this symplectic
structure of special geometry is related to $w_3=0$ in the differential
equation.

One can similarly discuss the properties of $\cD$ when in addition:
$$
w_4\ =\ 0\ .\eqn\wfnull
$$
{}From \wW\ it is clear that this applies in particular if $W=const$. However,
$w_4(W)=0$ is a non-trivial differential equation that possesses also other
solutions than $W=const$. One might thus ask about the significance of general
solutions of $w_4(W)=0$ with non-constant superpotential.

\ni If $w_4=0$, eq.~\wdeq\ simplifies to
$$
\cD V =  \left(
\del^4  + w_2\del^2 + w_2'\del +
\coeff3{10}w_2'' + \coeff9{100} w_2^2\right) V  \, ,
\eqn\wfeq
$$
and the solutions are given by
$
 \left\{ \theta_1^3,  \theta_1^2 \theta_2,
  \theta_1 \theta_2^2 ,\theta_2^3 \right\} \Forsyth  \, .
$
Here, $\theta_{1,2}$ are the independent solutions of
the second--order equation,
$$
(\del^2  + \coeff1{10}w_2) \theta_{1,2} = 0                \, .
\eqn\eqtheta
$$
One easily determines a symplectic basis to be
$$
\eqalign{
X^0 = \theta_1^3\, ,  \qquad
F_0 = - \coeff16{(X^1)^3 \over (X^0)^2} + 2 c_{00} X^0
+ 2 c_{01} X^1\ , \cr
  X^1 = \theta_1^2 \theta_2\, , \qquad
 F_1 = \coeff12 {(X^1)^2 \over X^0} + 2 c_{01} X^0 + 2c_{11} X^1\, , }
\eqn\symbasis
$$
where $c_{AB}$ are arbitrary constants.
Using the homogeneity property $X^A F_A = 2 F$ or integrating $F_A$
we find
$$
F=  \coeff16{(X^1(z))^3 \over X^0(z)} + c_{AB} X^A X^B \, .
\eqn\FXrel
$$
{}From this $F$ we can compute (using \ksolW) the Yukawa coupling and find that
it is covariantly constant: $\hat D\,W=0$. For $c_{AB} =0$ \FXrel\ is the
$F$--function\foot {
Note that quadratic terms in $F$ are generally not determined
by the \pf\ equations. They contribute to the physical Yukawa couplings
via $e^K$. It appears that they can be interpreted as
perturbative $\sigma$--model corrections in Calabi--Yau compactifications
\CandB.}
corresponding to the homogeneous moduli space $SU(1,1)/U(1)$ (which
satisfies the stronger constraint $D\,W=0$). Moreover, it follows from the
inhomogeneous transformation behavior \wtrans\ of $w_2$ that one can always
find
a ``Schwarzian'' coordinate where $w_2$ vanishes, by solving a Schwarzian
differential equation $\{t;z\}=\coeff{1}{5} w_2(t)$. Then one has $\theta_1 =
1, \theta_2 = t$ and thus can take ($c_{AB}=0$)
$$
\V\ =\ (1,t,\shalf t^2,\coeff16t^3) \ , \qquad  F\ =\
\coeff16t^3 \ ,\qquad W\ =\ \del^3 F = 1 \, .
\eqn\wtzero
$$
It is clear that $t=X^1/X^0$ is precisely the special coordinate of eq.\
\specoo\ (note that the coincidence of special coordinates with Schwarzian
coordinates holds only if $w_4=0$). There is an analogous group action that
preserves the relationship between the solutions of \wfeq. This group is just
the
invariance group of the Schwarzian derivative, which is $SL(2,\IR)$:
$\theta'=\coeff{a\theta +b}{c\theta+d}\ ,ad-bc=1$. The action on the solutions
of \wfeq\ is easily found through the mapping $V=\theta^3$:
$$
M\ =\ \pmatrix{ {a^3} & {a^2} c & {{a {c^2}}/ 2} & {{{c^3}}/ 6} \cr 3
{a^2} b & 2 a b c + {a^2} d & {{b {c^2}}/ 2} + a c d & {{{c^2} d}/ 2}
\cr 6 a {b^2} & 2{b^2} c + 4 a b d & 2 b c d + {a
{d^2}} & c {d^2} \cr 6 {b^3} & 6 {b^2} d & 3 b {d^2} & {d^3}
\cr } \ ,
\eqn\sltwot
$$
which is part of $Sp(4,\IR)$ \FLT. Thus, the specific structure of the
solutions is unique up to such $SL(2)$ transformations.

Summarizing, the above means that if $w_4=0$, the situation for generic $w_2$
is reparametrization equivalent to $w_2=0$, in which case the solutions are
given by \wtzero. This corresponds to a cubic $F$-function and to constant
Yukawa coupling, $W$. In general coordinates where $w_2$ does not vanish, $W$
is not constant (but covariantly constant with respect to the holomorphic
connections).

Thus, for covariantly constant Yukawa couplings
the differential equation is essentially reduced to the
differential equation of a torus. This is similar to the situation for the
$K_3$ surface where the only non--trivial $W$-generator is $w_2$ \LSW.
The possibility of having non--trivial Yukawa couplings, or $w_4 \neq 0$, is
the new ingredient in special geometry. It reflects the possibility of
having instanton corrections to $W$. Specifically, it is easy to see from
\simpdeq\ that in special coordinates the solutions have the general structure
$$
V\ =\ (\,1\,,t\,,\shalf t^2+\cO(t^3)\,,\coeff16 t^3+\cO(t^4)\,)\ ,\eqn\solexp
$$
where the higher order ``instanton'' terms arise from a non-trivial $w_4$.
Thus, the invariant $w_4$ measures the deviation from $W=const$, which is the
large--radius limit of the \cy\ moduli space. One can actually check that the
contribution of a given rational curve of degree $k$ to the Yukawa couplings
corresponds to a covariantly constant $w_4$
generator. That is, from \wW\ one finds that in special coordinates:
$w_4(W=e^{kt}) = (const)k^4$ (see also Appendix B).

We now turn to another way of understanding the significance
of $w_3=0$. This will also allow us to introduce some concepts
which nicely generalize to multi--dimensional moduli spaces (section 3).

\sect{First--order equations}

Any fourth--order linear differential equation \oddeq\ is
equivalent to a first--order matrix equation \DS
$$
\Big[\,{\bfone}\del\ -\ \IGa\,\Big]\cdot \bv\ =\ 0, \eqn\matdeq
$$
(for a particular choice of the matrix $\IGa$) where $\bv$ is a $4\times 4$
matrix whose first row is $V$. A matrix of the form
$$
\IGa\ =\ \pmatrix{
\ast & 1 & 0 & 0 \cr
\ast & \ast  & 1 & 0 \cr
\ast & \ast & \ast & 1 \cr
\ast & \ast & \ast & \ast \cr}
\eqn\formga
$$
corresponds to a fourth--order operator with $a_4=1$ whereas $tr \IGa = 0$
leads
to $a_3=0$. However, $\cD$ is left invariant by local gauge transformations
acting as $\bv\to S^{-1}\cdot \bv$ and $\IGa\to S^{-1}\IGa S-S^{-1}\del S$,
where $S$ has the form
$$
S\ =\ \pmatrix{
1 & 0 & 0 & 0 \cr
\ast & 1  & 0 &0 \cr
\ast & \ast & 1& 0 \cr
\ast & \ast & \ast & 1\cr}\in N \subset SL(4)\ .\eqn\Strans
$$
This is just the usual matrix of lower triangular transformations generated by
a nilpotent subalgebra of $sl(4)$. The top row of $\bv$ corresponds to the
highest weight and thus is also $N$--invariant (the other rows of $\bv$ are
gauge dependent). That is, the solutions of \wdeq\ are completely invariant
under the local transformations \Strans.

Note also that the more general gauge transformations belonging
to a Borel subgroup $B$ of $SL(4)$, where
$$
S = \pmatrix{
\ast & 0 & 0 & 0 \cr
\ast & \ast  & 0 & 0 \cr
\ast & \ast & \ast & 0 \cr
\ast & \ast & \ast & \ast \cr}\ \in\ B\ ,
\eqn\Stildetrans
$$
do not leave $\cD$ invariant but induce $a_3 \neq 0$ and $a_4 \neq 1$. However,
this just corresponds to a rescaling of the solution $\V\to f(z)\V$ (and
corresponds to an irrelevant \K\ transformation in this context).

\ni Using the gauge freedom one can put the connection to:
$$\IGa\ =\ \IGa_w\ \equiv\ \pmatrix{
0  & 1 &  0 & 0 \cr
-\coeff3{10}w_2 & 0  & 1 & 0 \cr
-\shalf w_3 & -\coeff4{10}w_2 & 0 & 1\cr
 -w_4 &-\shalf w_3 &-\coeff3{10}w_2 & 0\cr}\ \in sl(4,R)\ .\eqn\wmat
$$
To understand this form, recall the well-known relationship\foot{We
thank R.Stora for discussions on this point.} between $W$-algebras and a
special, ``principally embedded'' $SL(2)$ subgroup $\cS$ \Kos\ of $G=SL(N)$ (in
fact, $G$ can be any simple Lie group). The generators of $\cS$ are
$$
J_-=\sum_{{simple\atop roots\ \a}} b_\a E_\a\ ,\ \ \
J_+=\sum_{{ simple\atop roots\ \a}} c_\a(b_\a) E_{-\a}\ ,\ \ \
J_0=\rho_G\cdot H\ ,\eqn\gendef
$$
where $b_\a$ are arbitrary non-zero constants, $c_\a$ depend on
the $b_\a$ in a certain way and $\rho_G$ is the Weyl vector.
An intriguing property \Kos\ of $\cS$ is
that the adjoint of any group $G$ decomposes under $\cS$ in a
very specific manner:
$$
adj(G)\to \bigoplus r_j\ ,\eqn\sdecomp
$$
where $r_j$ are representations of $SL(2)$ labelled by spin $j$,
and the values of $j$ that appear on the r.h.s. are
equal to the exponents of $G$. The exponents are just the degrees
of the independent Casimirs of $G$ minus one (for $SL(N)$, they
are equal to $1,2,\dots,N-1$).

Recalling that the Casimirs are one-to-one
to the $W$ generators associated with $G$, one easily sees that
the decomposition \sdecomp\ corresponds to
writing the connection \wmat\ in terms of $W$-generators;
more precisely, for an $N$--th order equation
related to $G=SL(N)$, the connection \wmat\ can be written as
\doubref\Dublin\DIZ:
$$
\IGa_w\ =\ J_--\sum_{m=1}^{N-1} w_{m+1} (J_+)^m\ ,\eqn\wdec
$$
where $J_\pm$ are the $SL(2)$ step generators \gendef\ (up to irrelevant
normalization of the $w_n$).

In our case\foot {The choice \wmat\ for $\IGa$ corresponds to an embedding
\gendef\ with $b_1=b_2=b_3=1, $ and $c_1=c_3=3/10, c_2=4/10$.} with $N=4$, the
decomposition \sdecomp\ of the adjoint of $SL(4)$ is given by $j=1,2,3$, which
corresponds to $w_2,w_3$ and $w_4$. We noticed above that $w_3\equiv0$ for
special geometry and this means that $\IGa_w$ belongs to a Lie algebra that
decomposes as $j=1,2$ under $\cS$. It follows that this Lie algebra is $sp(4)$.
Indeed, remembering that the algebra $sp(n)$ is spanned by matrices $A$ that
satisfy $A\,Q + Q\,A^T=0$, we can immediately see from \wmat\ that
$$
\IGa_w\ \in\ sp(4)\ \qquad \ \ \longleftrightarrow\ \ \qquad \
  w_3\ \equiv\ 0\  .\eqn\spexpl
$$
Above, the symplectic metric $Q$ is taken as in \Symp.

We chose the gauge in \wmat\ such that the symplectic structure is manifest.
General gauge transformations conjugate the embedding of $sp(4)$ in $sl(4)$,
and in general gauges the fact that $\IGa_w\in sp(4)$ is not obvious.
The invariant way to express this fact is to state that $w_3=0$ in
the gauge invariant scalar equation.

Similarly, if in addition  $w_4=0$ (which corresponds to a covariantly
constant Yukawa coupling),
$\IGa_w$ further reduces to an $SL(2)$ connection.
This $SL(2)$ is identical to the principal $SL(2)$ subgroup, $\cS$,
since according to \wdec\ the entries labelled by $w_2$ and $1$ in
\wmat\ are directly given by the $\cS$ generators $J_+$ and $J_-$.
It consists precisely of the transformations \sltwot\ that
preserve the non--trivial relationship between the solutions.

\chap{Differential equations for arbitrary many moduli}
\sect{Holomorphic Picard-Fuchs equations and special geometry}

In this section we generalize the previous analysis to many variables. The
basic identities of special geometry are given by the system \deqset.\foot{The
relevant formulas of special geometry are collected in appendix A.} We already
mentioned in the introduction that, assuming that $(C_\a)_{\b\c}$ is
invertible, these identities are equivalent to
$$
D_\a D_\b ({C^{{-1}{\shat\c}}})^{\rho\sigma}D_{\shat\c}D_\s V\ =\ 0\ ,
\eqn\pfs
$$
where $\hat\c$ is a priori not summed over.

Since the solution vector $V\equiv (X_A(z),F^A(X))$ is holomorphic, we expect
that the non--holomorphic pieces in \pfs\ that come from the connections in $D$
cancel, so that \pfs\ is effectively a purely holomorphic identity. We will
prove below that this is indeed the case by showing that $V$ also satisfies
manifestly holomorphic identities that are equivalent to \pfs. These equations
contain only the holomorphic connections $\hat\Gamma$ and $\del\hat K$
(defined in Appendix A).

Let us choose special coordinates $t^a=X^a/X^0$ and the \K\ gauge $X^0=1$, and
consider the following set of equations:
$$
\eqalign{
\del_a V\ &=\ V_a \cr
\del_a V_b\ &=\ W_{abc} V^c \cr
\del_a V^b\ &=\ \delta_a^b V^0  \cr
\del_a V^0\ &=\ 0\ ,}
\eqn\spec
$$
where $(V,V_a,V^a,V^0)$ are all holomorphic and $W_{abc}$ are the Yukawa
couplings in special coordinates. The last two equations of \spec\ give
$$
V^0\equiv(1,0)\ ,\qquad \qquad V^a\equiv (t^a,\quad 1,\quad 0)\ ,
\eqn\last
$$
while the first two are solved by setting
$$
\eqalign{
V\equiv\ &(1,\quad t^a,\quad \del_a \cF,\quad t^a\del_a \cF-2 \cF)\ ,\cr
V_a\equiv\ & (0,\quad \delta_a^b,\quad \del_a\del_b \cF,\quad
t^b\del_a\del_b\cF
-\del_a \cF)\ .   }
\eqn\fa
$$
The holomorphic function $\cF$ is defined in eq.~(A.33) and satisfies   (in
special coordinates)
$$
\del_a\del_b\del_c \cF= W_{abc} \  .
\eqn\Wabc
$$
This identity is the only non--trivial input in solving
the differential equations. The system \spec\ can also be written in matrix
form,
$$
\eqalign{
(\,\bfone\del_a -  \IC_a\,)\,\bv\ &=\ 0\ ,\cr
\IC_a\ &=\ \pmatrix{0 & \delta_a^c & 0 &0\cr
0 & 0 & W_{abc} & 0 \cr
0 & 0 & 0 & \delta_a^b \cr
0 & 0 & 0 & 0}\ ,\cr}
\eqn\charg
$$
and from the above we see that this is solved by
the following $(2n+2)\times(2n+2)$--dimensional  matrix:
$$
\bv=
 \pmatrix{V \cr V_b \cr V^b \cr V^0} \ = \
\pmatrix{1 & t^a &\del_a \cF & t^a\del_a \cF-2\cF \cr
0 & \delta_b^a & \del_a\del_b \cF & t^a\del_a\del_b \cF-\del_b \cF \cr
0 & 0 & \delta_a^b & t^b \cr
0 & 0 & 0 & 1}\ .
\eqn\fmat
$$
{}From eqs.~\spec, (A.17)\ we can infer the transformation properties of $\bv$
under coordinate and K\"ahler transformation and thus it is straightforward to
write down the covariant and holomorphic version of eqs.~\spec:
$$
\eqalign{
\hat D_\a V\ &=\ V_\a \cr
\hat D_\a V_\b\ &=\ W_{\a\b\c} V^\c  \cr
\hat D_\a V^\b\ &=\ \delta^\b_\a V^0 \cr
\hat D_\a V^0\ &=\ 0 \ .}\
\eqn\holfp
$$
where $\hat D$ is defined in eq.~(A.29) and contains the holomorphic
connections given in eq.~(A.28). This system can also be written as
$$
\big({\bfone}\del_\a - \IGa_\a\big)\,\bv\ \equiv\ 0\ ,      \quad\quad
 \bv\ =\ \pmatrix{V \cr V_\b \cr V^\b \cr V^0}\ ,
\eqn\firstorder
$$
which contains the holomorphic ``connection''
$$
\IGa_\a\ =\
\pmatrix{
-\del_\a\Kh & \d_\a^\c & 0 & 0 \cr
0 & (\Gammah_\a-\del_\a\Kh\bfone)_{\b}^\c & (W_\a)_{\c\b} & 0\cr
0 & 0 & (\del_\a\Kh\bfone-\Gammah_\a)^\b_{\c} & \d^\b_\a \cr
0 & 0 & 0 & \del_\a\Kh\cr}\ .
\eqn\gammahat
$$
The general solution of \firstorder\ is just the covariant version of
eq.~\fmat\ and thus corresponds to the columns of the matrix
$$
\bv=\pmatrix{X^0 & X^a & X^0~e^\a_a \del_\a \cF  &
 X^a~ e^\a_a\del_\a \cF-2 \cF X^0\cr
0 & X^0~ e^a_\b & X^0 e^\a_a~ \hat D_\a~\del_\b \cF &
 X^a~ e^\a_a~ \hat D_\a~\del
_\b\cF-X^0~\del_\b \cF\cr
0 & 0 & (X^0)^{-1}~ e^\a_a & (X^0)^{-2}~ X^a~ e^\a_a\cr
0 & 0 & 0 &(X^0)^{-1}}\ .
\eqn\solu
$$
Here $e_\alpha^a = \del_\alpha t^a (z)$ which satisfies
$\hat{D}_\beta e_\alpha^a = 0$. Furthermore, in arbitrary coordinates
$\cF$ is \K\ invariant and obeys
$$
\hat D_\a \hat D_\b \hat D_\c{\cF} =(X^0)^{-2} W_{\a\b\c}\ \qquad \ .
\eqn\prepo
$$
The system \holfp\ implies the following
manifestly holomorphic equation for $V$:
$$
{\hat D}_\a {\hat D} _\b(W^{-1})^{{\hat \c}
\rho\sigma}{\hat D}_{\hat \c} {\hat D}_\sigma V =0.
\eqn\fp
$$
Using eq.~(A.33) one checks that the first row of \solu\ indeed coincides with
$V\equiv (X^A,F_A)$. We conclude, therefore, that eq.~\fp\ is the same as
eq.~\pfs, except that it is written in a manifestly holomorphic way.

 As for one variable, the correspondence between
 eq.~\fp\  and the linear system \firstorder\ is not unique.
Indeed, \fp\ is invariant under gauge transformations (up to \K\
transformations) acting on $\bf V$ and $\IGa$ via
$$
S\ =\ \pmatrix{ \ast_{1\times 1} & {\bf 0} & {\bf 0} & 0\cr
               \ast & \ast_{n\times n} & {\bf 0} & {\bf 0} \cr
               \ast & \ast & \ast_{n\times n} & {\bf 0} \cr
                \ast & \ast & \ast & \ast_{1 \times 1}}\ \in B,
\eqn\gaugetrans
$$
which belong to a Borel subgroup $B$  of $SL(2n+2,\IC)$.

It is easy to check that for one variable, the connection
$\IGa$ in \gammahat\ can be gauge transformed to the form \wmat\ that
displays the $W$-generators. More precisely, under a symplectic transformation
$$
S\ =\ {\rm diag}\big( W^{-1/2}, W^{-1/2}, W^{1/2}, W^{1/2} \big)
\eqn\gaugetra
$$
the connection $\IGa$ takes the form
$$
\IGa \ =\ \pmatrix{-\del{\tilde K} & 1 & 0 & 0 \cr
                      0 & {\hat \Gamma}-{\del {\tilde K}} & 1 & 0 \cr
                      0 & 0 & -{\hat \Gamma} + \del{\tilde K} & 1 \cr
                      0 & 0 & 0 & \del {\tilde K} }
\eqn\tildekappa
$$
where $\tilde K = \hat K +\coeff{1}{2} {\ln} W = - {\ln}
 (X^0 W^{-1/2}) $. To bring further
${\tilde K}$ to the gauge \wmat\ one obviously needs an additional
$Sp(4)$ transformation that belongs to the nilpotent subgroup $N$.
This transition from \tildekappa\ to \wmat\ is nothing but
a Miura transformation \DS.

We have seen in section 2 that the Picard-Fuchs equation for one variable can
invariantly be characterized by the vanishing of classical $W$-generators. The
vanishing of $w_3$ was related to $\IGa_w\in sp(4)$. For many variables, we do
not know how to characterize the differential equation \fp\ in terms of
covariant quantities like $w_n$. But in analogy to the one--variable equation,
we expect that the statement that corresponds to $w_3=0$ is just that $\IGa_\a
\in sp(2n+2)$ in \gammahat. Indeed, the gauge in which we wrote \gammahat\ is
manifestly symplectic: one easily verifies that $Q\IGa =(Q\IGa)^T$, where $Q$
is the symplectic metric given in \Symp.

More generally, we expect that a multi-variable equation can invariantly be
characterized by the subalgebra $g\subset sp(2n+2)$ in which the set
of connections actually takes values, for given $W_{\a\b\c}$
(just like for $n=1$ where the additional vanishing of $w_4$ implies that
$\IGa_w\in sl(2)$). For large $n$, there exists obviously a large
number of distinct possible subgroups. (Note that it is in general
not easy to determine $g$, as the embedding in $sp(2n+2)$ is gauge
dependent and thus not always obvious.
One is missing a gauge--invariant criterion for many variables,
in analogy to the vanishing of certain $W$-generators for one variable.)

The solution vectors can accordingly be viewed as representations of
$Sp(2n+2)$ (or of some subgroup). The set of solution vectors when
written as a matrix $\bv$ can always be chosen in a way such that this matrix
becomes a group element, by multiplying $\bv$ with an appropriate constant
matrix from the right. One can easily check that our choice of solution matrix
\solu\ is indeed symplectic with respect to the metric \Symp. In this way, one
can regard $\bv$ as a vielbein $V_\hata^\hatA$ with a well-defined symplectic
action on both indices ($\hatA,\hata=1,\dots,2n+2$).\foot{ The index $\hatA$
corresponds to a symplectic basis of the Hodge bundle $\cal H$ and the index
$\hata$ to the flat bundle $\cE$ defined in ref.~\Strom.} Under coordinate and
\K\ transformations $z\rightarrow {\tilde z}(z), \, K\rightarrow K+f(z) +{\bar
f}(\bar z)$, the matrix $\bv$ transforms as follows:
$$
{V_\hatb}^\hatA ({\tilde z})={S_\hatb^{-1}}^{\hata} (z) {V_\hata}^\hatB (z)
 {M_\hatB}^\hatA
\eqn\transform
$$
where $S$ is the symplectic block diagonal matrix
$$
S  =  \pmatrix{ e^{-f} & {\bf 0} & {\bf 0} & { 0}  \cr
                     {\bf 0} & e^{-f} \xi^{-1} & {\bf 0} & {\bf 0} \cr
                     {\bf 0} & {\bf 0 } & e^{ f} \xi & {\bf 0} \cr
                       0  & {\bf 0} & {\bf 0} & e^{ f} }\ \in B\ ,
\eqn\lambdamatrix
$$
with $\xi\equiv\xi^\a_\b=\del \tilde z^\a/\del z^\b$.
Furthermore, $M$ is a constant matrix that can always be taken as an element of
$Sp(2n+2)$. One easily infers from \gammahat\ that these transformations
are nothing but  gauge transformations of the holomorphic connections $\del
\Kh$ and $\hat \Gamma$:
$$
\eqalign{
\del_\a\Kh\ &\longrightarrow\  \del_\a \Kh + \del_\a \,f\cr
\Gammah_\a\ \ &\longrightarrow\  \xi^{-1}\Gammah_\a\xi + \del_\a \ln\,\,\xi\
.\cr}\eqn\gaugetransf
$$
This point of view allows us to also understand how global $Sp(2n+2)$
transformations acting on the index $\hatA$ induce local frame rotations acting
on the index $\hat\a$: the local rotations are induced by the requirement that
$\IGa_\a$ stays in the gauge \gammahat. More explicitly, symplectic
transformations: $\tilde{\bv}=\bv\cdot M$, which act in particular on the
solution vector as
$$
\eqalign{
\big(\tilde X^A , \tilde F_A(\tilde X^A)\big)\ &=\
\big(X^A,  F_A(X^A)\big)\cdot M\cr
M\ &=\ \pmatrix{ A & C \cr B & D\cr}\ \in Sp(2n+2)\ ,\cr}
\eqn\spdef
$$
induce the following reparametrizations of special coordinates:
$$
\tilde t^a\ =\ {A^a_BX^B+B^{aB}F_B\over A^0_BX^B+B^{0B}F_B}(t)
\eqn\repa
$$
These reparametrizations
induce local, compensation gauge transformations \gaugetransf\ with
$f=\Tr(\ln\xi)$ and $\xi=\del \tilde t^a/\del t^b$.

Note that the transformations \lambdamatrix\ belong to the part
of the Borel gauge group \gaugetrans\ that is not fixed by the gauge choice
\gammahat; that is, they lie in (the complexification of) the maximal compact
subgroup $U(n)\times U(1)$ of $Sp(2n+2)$. This implies that the group
element $\bv$ can be thought of as an element of $G/H$, where $G\subset
Sp(2n+2)$ and $H\subset U(n)\times U(1)$. More specifically, one can
decompose \doubref\Strom\CeVa
$$
\IGa_\a\ =\  \IGam_\a + \IC_\a\ ,
\eqn\ac
$$
where the diagonal part, $\IGam_\a$, consists of the connections $\hat\Gamma$
and $\del\hat K$ (which are flattened by special coordinates $t^a=X^a\, ,
X^0=1$). Furthermore, $\IC_\a$ is the covariant version
of  \charg\ and generates an Abelian,
$n$-dimensional subalgebra of $sp(2n+2)$ that is nilpotent of order three:
$\IC_\a \IC_\b \IC_\c\IC_\d =0$. Thus, $G$ is determined by the subalgebra of
$sp(2n+2)$ in which $\IC_\a$ takes values, and $H$ is determined by the
subgroup of $U(n)\times U(1)$ that is gauged by $\IGam_\a$ \CeVa.

More precisely, $\bv$ is an element of $G^c/B$ (which is, essentially,
isomorphic to $G/H$), where $G^c$ is the complexification of $G$ and $B$ the
Borel subgroup \gaugetrans, which contains the complexification of $H$. From
this viewpoint one can easily make contact to supersymmetric $\sigma$-models on
moduli spaces. According to \WL, \K\ potentials for homogeneous \K\ manifolds
$G/H$ can be written in terms of holomorphic CCWZ type coset representatives
$L\in G^c/B$ as arbitrary functions of $K_0\equiv vLQL^\dagger v^\dagger$.
Here, $L$ transforms under global $G$ transformations as: $L(z)\,g=S(z)L(z)$,
where $g\in G$ and $S\in B$. Furthermore, $Q$ is the metric of $G$ and $v$
denotes an isotropy vector, which is left invariant under $S$ (up to a $U(1)$
factor, which corresponds to \K\ transformations). Note that $K_0$ is
manifestly invariant under global $G$ and under local $S$ transformations
(except for the \K\ transformations). Taking for $Q$ the symplectic metric
\Symp, $v=(1,0,0\dots,0)$ and $L=\bv$, the logarithm of $(-i)K_0$ gives
precisely the \K\ potential \ksolK\ of special geometry: $v\bv Q\bv^\dagger
v^\dagger=-(X^A \bar F_{A}-\bar X^{A}F_A)$.

In the generic case, $G/H=Sp(2n+2)/U(n)\times U(1)$, but the moduli space
in which $\bv(z)$ actually takes values is a complicated subvariety of this
space. However, there are special cases where $G$ and $H$ are effectively
smaller subgroups; one example are the theories with cubic $F$--function
where the moduli spaces are directly given by $G/H$. For instance for $n=1$,
the generic moduli space is some complicated one-dimensional submanifold of
$Sp(4,\IR)\over U(1)^2$ whose complex dimension is four. But for constant
coupling $W$ (and for $c_{AB}=0$ in \FXrel), the moduli space in which $\bv$
takes values is the one--dimensional submanifold $G/H={SL(2,\IR)\over U(1)}$.
The special geometry of cubic $F$--functions is further discussed below in
Appendix C.

\sect{Non--holomorphic \pf\ equations}

In this section we establish the relationship between the first--order
systems \deqset\ and \holfp. Let us first note that the gauge group
\gaugetrans\ can also be extended to non--holomorphic gauge
transformations ${\cal S} = {\cal S} (z,{\bar z})$ that leave $V$ invariant.
The point is that eqs.~\deqset\ and \holfp\ are precisely related by such a
non--holomorphic gauge transformation. Thus the non--holomorphicity of
\cwrelation\ and \deqset\ can be viewed as gauge artifact and this reflects
the fact that all quantities in special geometry are determined entirely in
terms of holomorphic quantities.

More specifically, one can rewrite the non--holomorphic system \deqset\ in
first--order form
$$
\cD_\a\,{\bf U}\ \equiv\ (\bfone{\del_\a} - {\cA_\a}) {\bf U}\ =\ 0\ ,
\eqn\linearsi
$$
where ${\bf U}=(V,U_\a,\bar{U}_{\bar \a},\bar{V})^T$ and
$$
\cA_\a\ =\ \pmatrix{
-\del_\a K & \delta_\a^\b & 0 & 0 \cr
0 &- \delta_\c^\b \del_\a K + \Gamma _{\c\a}^\b &
- i C_{\a\b\c}g^{\c \bar{\c}}& 0\cr
0 & 0 & 0 & g_{\a\bar{\b}} \cr
 0 & 0 & 0 & 0 } \ .
\eqn\gammanonhol
$$
In addition ${\bf U}$ also satisfies
$$
\cD_{\bar\a}\,{\bf U}\ \equiv\ (\bfone{\del_{\bar\a}} - {\cA_{\bar\a}})
{\bf U}\ =\ 0\ ,
\eqn\complexco
$$
where
$$
\cA_{\bar\a}\ = \ \pmatrix{
0 & 0 & 0 & 0 \cr
g_{\bar{\a}\b} & 0 & 0 & 0 \cr
0 & i C_{\bar\a\bar\b\bar\c} g^{\bar\c \c }
& -\delta _{\bar \c} ^ {\bar \b} \del_{\bar
\a} K +\Gamma_{\bar\c\bar\a}^{\bar \b} & 0 \cr
0 & 0 & \delta_{\bar{\a}}^{\bar{\b}} & -\del_{\bar\a} K }\ .
\eqn\gacomplexco
$$
It is easy to verify that as a consequence of \cwrelation\ the connections
$\cA_\a$ and ${\cA}_{\bar\a}$ have vanishing curvature \Strom\foot{Vice versa
\Strom, one can start from a covariantly constant basis $(V, U_\a, U_{\bar\a},
\Vb)$ of a flat $Sp(2n+2,\IR)$ vector bundle $\cE$ with connection $\cA$ and
derive the fundamental identities \linearsi\ and \complexco\ of special
geometry.}:
$$
[\,\cD_\a,\cD_\b\,]\ =\ [\,\cD_{\bar\a},\cD_{\bar\b}\,]\ =\
[\,\cD_\a,\cD_{\bar\b}\,]\ =\ 0\ .
\eqn\zerocu
$$
It follows that via non--holomorphic transformations ${\cal S}(z,{\bar z})$
that leave $V$ invariant (${\cal S} V=V$), one can gauge away $\cA_{\bar\a}$
and make $\cA_\a$ purely holomorphic. As a consequence of eq.~\zerocu\ one can
go to a gauge where
$$
{\cA}_{\bar\a}=
{\cal S} \del_{\bar\a} {\cal S}^{-1} \, .
\eqn\puregauge
$$
This implies
$$
\del_{\bar\a} ({\cal S} {\bf U})\ =\ 0\qquad {\rm and}
\qquad \del_{\bar\a}\big[
 {\cal S}{\cA}_\a {\cal S}^{-1} - {\cal S} \del_\a {\cal S}^{-1}\big]\
=\ 0\ ,
\eqn\gautra
$$
so that the non--holomorphic system \linearsi\ becomes the holomorphic system
\firstorder\ with
$$
\IGa\ =\ {\cal S}\cA_\a {\cal S}^{-1} - {\cal S} \del_\a {\cal S}^{-1}\
,\qquad\qquad {\bv}= {\cal S}\, {\bf U}\ ,
\eqn\sese
$$
which displays a residual gauge symmetry of holomorphic transformations. Of
course, one could also have chosen gauge transformations $\bar {\cal S}$ that
leave the lowest row of $\bv$, ie. $\Vb$, invariant; in that case one would
have produced a purely anti--holomorphic connection $A_{\bar\a}$. The point is
that there is no invariant subspace with respect to both $\cal S$ and $\bar
{\cal S}$, so that the connection cannot be completely flattened.

\sect{Singular Picard-Fuchs systems}

In the previous considerations we have assumed that the
 matrices $(W_\a)_{\b\c}\equiv W_{\a\b\c}$ are invertible for all $\a$.
It is interesting to find the implications of degenerate
 fourth--order partial differential equations \fp.

i) The simplest situation is when $W_{\a\b\c}\equiv 0$.
 Then, from \holfp\ or alternatively from \deqset\ one can see
that the Picard-Fuchs identities become of second order
$$
\hat{D}_\a \hat{D}_\b V=0 \, .
\eqn\secondor
$$
 In special coordinates we get
$$
\del_a\del_b V=0
\eqn\secspe
$$
with solutions $(1\, ,t^a)$. This corresponds to $\cF=(t^a)^2-\coeff{1}{2}$,
and implies that the symplectic connection \gammahat\ becomes block diagonal in
two $(n+1)^2$ blocks. The matrices $\IC_a$ are nilpotent of order two ($\IC_a
\IC_b=0$), and the solution matrix is given by
$$
\bv=e^{t^a C_a} \ .
\eqn\constyuk
$$
The moduli space is (locally) $G/H= U(1,n)/U(1)\times U(n)$ with the embedding
$2n+2=(n+1)\oplus (n+1)$ of $U(1,n)$ in $Sp(2n+2)$.

ii) We now consider the situation in which $W_{\a\b\c}$ does not vanish but
is degenerate. This is best discussed in special coordinates where $\IGam_\a=0$
and $\IGa_\a=\IC_\a$. Let us first consider $W_{ijk}=0$ for
 some subset  of indices $i,j,k,$ and also  $W_{i ab}=0,W_{ijb}=0$. Then,
assuming that the remaining couplings $W_{abc}$
  give rise to an invertible matrix $(W^a)^{bc}$, we have two sets
of decoupled equations
$$\eqalign{
\del_i\del_j V\ &=0\cr
\del_a\del_b(W^{-1})^{cde}\del_c\del_d V\ &=0}
\eqn\somezero
$$
with solutions given by the prepotential
$$
\cF(t^i,t^a)=c+(t^i)^2+\cF(t^a)     \ .
\eqn\twovar
$$
To write these equations in arbitrary coordinates we note
$$
\hat{D}_\a V_\b=W_{\a\b\c}V^\c\equiv
 \del_\a t^A\del_\b t^B\del_\c t^C W_{ABC} V^\c
\eqn\arbit
$$
(where $A, B, C$ here stands for either $a, b, c$ or $i, j, k$).
Equivalently, multiplying by the inverse vielbeins
 $e_A^\a\equiv(e_\a^A)^{-1}=(\del_\a t^A)^{-1}$ one gets:
$$
e_A^\a e_B^\b \hat{D}_\a V_\b=W_{ABC} V^\c e_\c^C
\eqn\anysys
$$
Supposing $W_{ABC}=W_{abc}, W_{ijk}=W_{aij}=W_{abj}=0$, we get
$$
\eqalign{
e_i^\a e_j^\b \hat{D}_\a \hat{D}_\b V\ &= 0 \ ,\cr
\hat{D}_\rho \hat{D}_\mu
(W^{-1})^{abc} e_a^\a e_b^\b \hat{D}_\a \hat{D}_\b V\ &=0\ .}
\eqn\mixed
$$
Using $\hat{D}_\a e_\b^a=0$, the last equation can also be written as
$$
e_d^\rho e_e^\mu e_a ^\a e_b ^\b
\hat{D}_\rho \hat{D}_\mu (W^{-1})^{abc} \hat{D}_\a \hat{D}_\b V=0 \, .
\eqn\also
$$
In \anysys\ and \mixed\ all moduli coordinates appear, but the
structure of the equations is such that they indeed become
decoupled by making a coordinate transformation.
 The coordinate independent statement on the Yukawa couplings is
$$
e^\a_i e^\b _j e ^\c_k W_{\a\b\c}=0
\eqn\yukaw
$$
for a subset $(i,j,k\subset A,B,C);\ A,B,C=1,\cdots,n $.

iii) There are two more special cases of interest. One corresponds to two
non-vanishing  Yukawa couplings for different sets of indices
$W_{ijk}\neq  0, W_{abc}\neq 0$ with $ W_{iab}=W_{ija}=0$.
 In this case one gets two sets of fourth--order equations of the type
\also\ .

iv) The other case is when $W_{ijk}=0, W_{ija}=0$ but $W_{iab}\neq 0$.
Here the matrix $W_i$ is invertible in the subblock $(a,b)$
and the matrix $W_{a,BC}$ is fully invertible. The prepotential,
 in special coordinates, is of the form
$$
\cF(t^i,t^a)=C_{ij}t^i t^j+ h_i(t^a)t^i+h(t^a)\qquad\qquad C_{ij}={\rm
const.}
\eqn\prespe
$$
In this case we get three sets of decoupled differential equations
$$
\eqalign{
e_i^\a e_j^\b \hat{D}_\a \hat{D}_\b V\ &= 0\cr
e_a^\mu e_A^\rho \hat{D}_\a
\hat{D}_\b (W^{-1})^{aAB} \hat{D}_\mu \hat{D}_\rho V\ &= 0\cr
e_a^\rho e_i^\mu \hat{D}_\a
\hat{D}_\b (W^{-1})^{iab} \hat{D}_\rho \hat{D}_\mu V\ &= 0
 .}
\eqn\threesets
$$
The purpose of these exercises was to find the differential equations for
decoupled chiral rings. In special coordinates, this reflects in a simple
additive structure of $F$. On the other hand, the corresponding \K\ metrics
by no means have the structure of direct product manifolds, and are rather
quite complicated. This shows that special geometry is most easily
characterized by $F$ and not by the geometry of the underlying manifold.

\chap{Relation to Calabi-Yau manifolds and topological field theory}

The discussion of sections 2 and 3 has been completely general and without
any reference to Calabi-Yau moduli spaces or to more general $c=9$, $(2,2)$
superconformal field theories. In this section, we relate our discussion to the
special case
of Calabi-Yau manifolds and to topological \LG theories. We understand
that part of the material of this section is well-known (see, for example,
\refs{\Griff,\Strom,\CandA,\Cec,\CeVa,\BVar,\LSW}), but we think it is
important to give the precise relationship to special geometry. This
relationship is useful for practical computations.

We like first to review some properties of the period matrix
\doubref\Griff\CandA.
For those special geometries for which there exists an underlying
Calabi-Yau space $\cM$, the sections $V,U_\a,\bar U_{\bar\b}$ and $\bar V$  in
the non-holomorphic system \deqset\ can be viewed as basis elements of the
third
real cohomology of $\cM$, that is, of
\def\hd#1#2{H_{\bar\del}^{(#1,#2)}(\cM,\IR)}
$\hd30,\hd21,\hd12$ and $\hd03$, respectively. Furthermore, the solutions
$(X^A, F_A)$ of the Picard-Fuchs equation \fp\ are just the periods of
the holomorphic three--form, $\Omega$ \refs{\Griff{,}\CandA{,}\CO}:
$$
X^A\ =\ \int_{\gamma_A}\Omega\ ,\ \qquad F_B\ =\ \int_{\gamma_B}\Omega\
\eqn\Omperiods
$$
($A,B=0,\dots,n$ where $n\equiv h^{2,1}$).
Here, $\c_A, \c_B$ are the usual basis cycles of $H_3(\cM,\IR)$. More
generally, the complete
solution matrix ${V_\hata}^\hatA$ of
the first--order system \firstorder\ can be interpreted as the period matrix of
$\cM$,
$$
{\Pi_\hata}^\hatA\ =\ \pmatrix{
\int_{\gamma_\hatA}\Omega\cr
\int_{\gamma_\hatA}\chi_\a\cr
\int_{\gamma_\hatA}\bar \chi_{\bar\b}\cr
\int_{\gamma_\hatA}\bar \Omega\cr}\ ,\qquad\ \ \hata,\hatA=1,\dots,(2n+2)\ .
\eqn\pmatdef
$$
It is well-known \Griff\ that the period matrix is defined only up
to local gauge transformations,
$$
\Pi\ \sim\ S\,\Pi\ ,\ \qquad\ \
S = \pmatrix{
\ast & 0 & 0 & 0 \cr
\ast & \ast  & 0 & 0 \cr
\ast & \ast & \ast & 0 \cr
\ast & \ast & \ast & \ast \cr}\ \in B\ ,
\eqn\Strans
$$
and this is precisely the gauge symmetry \gaugetrans\ of the first--order
system.
Thus, we can represent the period matrix also in the holomorphic gauge \solu,
where the non-holomorphic sections $V,U_\a,\bar U_{\bar\a}, \bar V$
are replaced by holomorphic basis elements $(V,V_\a,V^\a,V^0)$ of $H^3$.

In addition, the period matrix is equivalent under conjugation by an integral
matrix, $\Lambda$: $\Pi\sim\Pi\Lambda$. These transformations $\Lambda\in
Sp(2n+2,\ZZ)$, which correspond to changes of integral homology bases, preserve
the symplectic bilinear intersection form $Q$ of $H_3(\cM,\ZZ)$, that is:
$\Lambda\,Q\,\Lambda^T=Q$ (the subset of these transformations that leave $F$
invariant up to redefinitions constitute the ``duality group''). This
intersection form is at the origin of the symplectic structure of the period
matrix. More precisely, denoting the $(n+1)\times(2n+2)$-dimensional submatrix
${\Pi_\ell}^{\hatA}$, $\ell=0,\dots,n$ by $\hat\Pi$, then one has in
general\foot{For $d$-dimensional complex manifolds, the invariance group of $Q$
is $Sp(b^d)$ if $d$ is even, and is equal to $SO(b^d_+,b^d_-)$ if $d$
is odd \Griff. As a consequence of this and according to our discussion in
section 2, it follows that in the differential equations for one variable one
necessarily has $w_n\equiv0$ for odd $n$ \CeVa. This generalizes $w_3=0$ for
Calabi-Yau spaces.}\Griff
$$
\hat\Pi\,Q\,\hat\Pi^T\ =\ 0\ .\eqn\bili
$$
This equation is analogous to Riemann's bilinear identity for period matrices
of 2d surfaces, and is satisfied for $\Pi\in Sp(2n+2)$. This is indeed
a general property of the solution matrix $\bv$ \solu\ in special
geometry\foot{Of course, not every solution matrix $\bv$ of special geometry
needs to correspond to a period matrix of some Calabi-Yau space;
this is a variant of the Schottky problem. Thus, special geometry is more
general than compactification on \cy\ manifolds.}.

On the other hand, in special geometry there is no intrinsic notion of homology
cycles. Rather, the symplectic structure arises from the
appearance of $Q$ in the \K\ potential,
$$
K = -\ln\langle\,\Omega\,|\,\bar\Omega\,\rangle\
=\ -\ln [V\,(-iQ)\, V^\dagger]\
,\eqn\kapot
$$
where $\langle\,x\,|\,y\,\rangle\equiv\int_\cM x\wedge y$ and $V$ is equal to
the first row of the symplectic matrix $\bv$. As we have seen in the previous
sections, the existence of such a solution matrix is guaranteed by the fact
that the connection in the first--order system \firstorder\ is symplectic. This
matrix equation, in turn, is an expression of $W_{\alpha\beta\gamma}=
\del_\alpha X^A \del_\beta X^B \del_\gamma X^C F_{ABC}$. Thus, in a sense, it
is this
identity that is responsible for the symplectic structure of special geometry.

Of particular physical interest are the Yukawa couplings,
$C_{\a\b\c}\ =\  e^K \WABC$.
According to \doubref\CandA\Strom, for a given \K\ potential \kapot\
the holomorphic sections $W_{\a\b\c}$ can be written as:
$$
\WABC\ =\ \langle\Omega|\del_\a\del_\b\del_\c\Omega\rangle\ .
\eqn\Wabcdef
$$
It is crucial to note that equations \kapot\ and \Wabcdef\ determine the gauge
of $\WABC$ in terms of the gauge of $K$, so that there is no ambiguity in the
physical couplings $C_{\a\b\c}$. One can easily check that the derivatives in
\Wabcdef\ can be replaced by covariant derivatives for free, reflecting the
fact that $\WABC$ is \K\ and reparametrization covariant. {}From the
first--order systems \linearsi\ or \firstorder\ it is clear that
differentiation by
$\del_\a$ on $H^3$ is equivalent to multiplication by the matrix $\IGa_\a$.
Thus, the holomorphic couplings can be written as a triple matrix product:
$$
\WABC\ =\ \big(\IGa_\a\IGa_\b\IGa_\c\big)_V^{\ V^0}\langle V|V^0\rangle\
.\eqn\wabcmat
$$
Considering the form of $\IGa_\a$ in either the non-holomorphic gauge
\gammanonhol\
or in the holomorphic gauge \gammahat, it is easy to see that eq.\ \wabcmat\ is
indeed identically satisfied.

Let us now discuss how $\WABC$ and $K$ can be computed explicitly. One method
is to evaluate the period integrals \Omperiods, using $X^AF_A=2F$ to obtain
$F$. This is how the Yukawa couplings for the quintic have been computed in
\CandB. In general, however, it seems to be easier to solve the Picard-Fuchs
differential equations \doubref\Ferr\LSW. These equations, though, just
represent identities ultimately expressing the fact that
$\WABC=\del_\a\del_\b\del_\c F$, and depend explicitly on the unknowns $\WABC$.
Thus, one needs additional information in order to pin down the explicit form
of these equations. This additional input makes use of the fact that the period
matrix can be represented in a very specific way.

To be more precise, consider first the
perturbed, quasi-homogeneous potential
$$
\W(x_i,\mu_\a)\ =\ \W_0(x_i) - \sum \mu_\a\,p_\a(x_i)\ ,\qquad\ \
\a=1,\dots,h^{2,1}\ ,
\eqn\suppot
$$
where $\W=0$ describes the Calabi-Yau manifold in question (for simplicity, we
restrict our discussion to hypersurfaces in weighted projective $4$-space; the
generalization is straightforward).
Above, $p_\a(x_i)$ denote the marginal operators (which are polynomials in the
homogeneous coordinates $x_i$), and the dimensionless moduli $\mu_\a$ are
certain functions of the flat coordinates $t^a$. As is well-known, $\W$ can be
viewed as the superpotential of a \LG theory that describes the underlying
\nex2 superconformal field theory \GVW, but this interpretation is not
necessary in the present context.

The non--trivial point is that the period matrix can be represented,
in a particular, holomorphic gauge as follows \Griff:
$$
{\Pi_\hata}^{\hatA}\ =\ \int_{\gamma_{\hatA}}\,{ \phi_\hata(x_i)\over
 \W^{\ell(\hata)}(x_i,\mu_\a)}\omega\ .
\eqn\resid
$$
Here, the homology cycle $\gamma_\hatA$ is a basis element of $H_3(\cM,\IR)$,
$\omega$ an appropriate volume form and $\ell(\hata)$ is determined by the
degree of the homogeneous polynomial $\phi_\hata(x_i)$ (so that the integrand
scales correctly). In general, these
$\phi_\hata(x_i)$ can be any basis of the local ring $\cR$ of $\W$, but we
restrict here only to those polynomials that represent the third cohomology of
the Calabi-Yau space. They generate a subring which we denote by
$\Rt$. More specifically, we choose $\Rt=\{\phi_\hata\}=\{1,p_\a,p^\b,\rho\}$,
where $p_\a$ are the marginal operators in \suppot, $\rho$ is the unique
top element of $\cR$, and $p^\b$ can be defined such that
$p_\a\, p^\b=\delta^\b_\a\rho$. Clearly, $\phi_\hata=1,p_\a,p^\b,\rho$ are
associated with differential forms belonging to
\def\hd#1#2{H^{(#1,#2)}}
$\hd30,\hd21,\hd12,\hd03$, respectively.

For example, for the quintic discussed in \CandB\ with
$\W=\sum_{i=1}^5 (x_i)^5 - \mu X$ (where $X\equiv x_1x_2x_3x_4x_5$),
the subring $\Rt$ consists of elements $\phi_\hata=X^{\hata}$,
$\hata=0,\dots,3$, which are associated with $\hd{3-\hata}\hata$, respectively.

The period matrix \resid\ identically
satisfies the following holomorphic, first--order ``Gau\ss-Manin'' system:
\def\per#1.#2{\int\!\coe#1.\W^{#2}.\omega}
$$
\Big[\, \bfone{\del\over\del \mu_\a}\ -\ \IGa_\a(\mu)\,\Big]\, \Pi\ =\ 0\ , \ \
 \ {\rm where}\ \ \
\Pi\ =\ \pmatrix{\per1.{}\cr\per p_\b.2\cr\per p^\c.3\cr\per\rho.4\cr}\ ,
\eqn\tildesyst
$$
and $\IGa_\a=\IGam_\a+\IC_\a$, with
\def\pn{\phantom{*}}
$$
\IGam_\a=\pmatrix{* & 0 & & \cr * &*& 0 &
\cr *&*&*&0\cr *&*&*&*\cr}\ , \qquad\ \ \
\IC_\a=\pmatrix{\pn & 1 & & \cr \pn &\pn& (W_\a^{(p)})_{\b\c} &
\cr \pn&\pn&\pn&1\cr \pn&\pn&\pn&\pn\cr} \ .
\eqn\gaussmanin
$$
This system can be seen as a gauge and coordinate transform of the holomorphic
special geometry system \firstorder\ (and also of the non-holomorphic system
\gammanonhol). The matrices $\IC_\a$
are the structure constants of the subring
$\Rt$, and the couplings $(W_\a^{(p)})_{\b\c}$ are determined by simple
polynomial multiplication modulo the vanishing relations, ie., by $p_\a
p_\b=W^{(p)}_{\a\b\c}p^\c$ mod $\nabla \W$. The crucial point is that also the
components of $\IGam_\a$ can be easily computed\foot{They arise from the
$\nabla \W$ piece, by partial integration. Note that the above expansion of
$p_\a p_\b$ is in general by no means unique, reflecting the gauge freedom in
\tildesyst.} directly from $\W$ (this is explained in detail in
\doubref\Ferr\LSW).

One way to solve the system \gaussmanin\ is to go to flat coordinates
$t^a\equiv X^a/X^0$ where the Gau\ss-Manin connection vanishes. As was shown in
detail in \LSW, imposing this condition gives a differential equation that
determines explicitly the dependence of the \LG couplings $\mu_\a$ on the
$t^a$. (The precise form of this complicated, non-linear differential equation
is not important here and can be inferred from \LSW.) In such flat coordinates
and in an appropriate gauge, the first--order system takes the form \charg.
In going to \charg, we implicitly compute eq.\ \wabcmat
\def\da#1#2{ {\del \mu_{#2}\over\del t^{#1}}}
$$
\wabc(t)\ =\ W^{(p)}_{\a\b\c}\,\da a\a\da b\b \da c\c\,\langle\, V\,|\,
V^0\,\rangle\ ,
\eqn\wabcrho
$$
where $V=\int\coeff1\W\omega$ and $\V^0=\int\coeff\rho{\W^4}\omega$.
One can view \wabcrho\ as a change from a topological basis (with indices
$\a,\b,\c$) to a flat basis (with indices $a,b,c$), and
$\langle\, V\,|\,V^0\,\rangle$ as a change of \K\ gauge.
It can be inferred from \CandA\ that
$$
\langle\, V\,|\, V^0\,\rangle\ =\
\int_{\Gamma_5}{\rho\over \del_1\W\dots\del_5\W}d^5\!x\ \equiv\
\tft\rho\ ,\eqn\rhovac
$$
where $\Gamma_5$ is the direct product of five one-dimensional contours that
wind around the five curves $\del_i\W=0$, and $\tft{\ }$ denotes the
Grothendieck residue \doubref\DVV\Vafa. It has the property: $\tft H=1$ (up to
a constant), where $H$ is the Hessian of the superpotential, that is,
$H\equiv\det[\del_{x_i}\del_{x_j}\W(x_k)]$. In general, $\rho$ and $H$ differ
by some holomorphic function and vanishing relations, $\rho=f_H H$ mod
$\nabla\W$, so that
$$
\langle\, V\,|\, V^0\,\rangle\ =\ \tft\rho\ =\ f_H\ .
\eqn\Veval
$$
For example, for the quintic: $W^{(p)}_{111}=1$ and $f_H\sim\coe1.1-\mu^5.$.
It follows that $W_{111}\sim\coe (\mu')^3.1-\mu^5.$, which is the result of
\CandB\ (in a particular gauge).

In topological \LG theory \doubref\DVV\Vafa\ one considers three--point
correlators:
$$
\tft{\Phi_a\Phi_b\Phi_c}\ =\ \wabc^{(top)}\,\tft H\ ,
\eqn\topcorr
$$
where $\tft{\ }$ has exactly the same meaning as in \rhovac\ and where the flat
fields are defined \DVV\ by: $\Phi_a(x_i,t^b)=-\coe\del.\del
t^a.\W(x_i,\mu(t^b))$. Referring back to the form of the superpotential
\suppot, one quickly sees that one indeed computes here absolutely the same
thing as in \wabcrho, that is\foot{A similar result was obtained in \GiSm, but
the precise relation of $\wabc^{(top)}$ to special geometry remained unclear.}:
$\wabc^{(top)}\equiv \wabc$, and the \K\ potential in the corresponding gauge
is: $K=-\log\langle V|\bar V\rangle$.

This is of course as expected, since also $\wabc^{(top)}$ can be represented as
a triple derivative of some function $F$ \DVV. One might think that this fact
already proves the equality of $\wabc^{(top)}$ and $\wabc$ of special geometry,
defined in \Wabcdef. However in \DKL\ it was shown that there generally exist
at least two different coordinate systems where
$W_{\a\b\c}=\del_\a\del_\b\del_\c F$ with two different and inequivalent \K\
potentials that solve the defining equations \deqset\ of special geometry.
Given this potential ambiguity, we feel, therefore, more comfortable displaying
explicitly the relationship between the couplings $\wabc^{(top)}$ computed in
topological field theory on the one hand, and $\wabc$ and $K$ of special
geometry on the other.

Note that rescaling $\W\to e^{f(t)} \W$ gives an equivalent superpotential.
{}From \resid\ we see that this just amounts to a \K\ transformation,
$\Omega\to e^{-f}\Omega$. Therefore a better, covariant way to specify flat
fields is
$$
\eqalign{
\Phi_\a(x_i,t^b)\ =\ - \hat\nabla_\a\,\W(x_i,t^b)\ &\equiv\
- \big[\,\del_\a -
\del_\a\Kh\,]\,\W(x_i,t^b)\ ,\cr &{\rm where}\ \ \del_\a\Kh
\equiv -
\del_\a\log\,X^0\ ,\cr}
\eqn\covder
$$
such that $\Phi_\a\to e^{f}\Phi_\a$.
 This transformation behavior is actually required for consistency of
\topcorr\ as $\wabc\to e^{-2f}\wabc$ and $H\to e^{5f}H$.

\chap{Outlook}

The investigation carried out in this paper leads to a number of open
questions. We saw that in a one--dimensional parameter space the \pf\ equation
is characterized by $W$--generators. We believe that it is presently not known
what the analogue of $W$--generators in higher dimensional parameter spaces
would
be. It seems interesting to find these ``invariants'' of partial differential
equations and their generalized $W$--algebra. Furthermore, we have seen
that for one variable, instanton corrections to the Yukawa couplings are
related to the $w_4$--generator. Thus, there might be a generic relation
between the geometry of the $W$--algebra and instantons on \cy\ manifolds. The
multi--variable analogue might elucidate the general structure of instanton
corrections to the effective action; it might be instructive to consider
a higher dimensional example explicitly.

Another topic we did not touch upon in this paper is the quantum duality
symmetry. For \cy\ compactification the duality group is related to the
monodromy group of the differential equation \nrf{\Font}
\refs{\CandB{,}\LSW{,}\DM{,}\Font}, which in turn depends on the zeros and
poles of
the Yukawa couplings. We are confident that the covariant formulation \fp\ of
the \pf\
equations presented in this paper will help to understand the general
properties of the duality symmetry.

\ack

We would like to thank P.\ Candelas, B.\ de Wit, C.\ Itzykson, A.\ Klemm, D.\
L\"ust, R.\ Stora
and S.\ Theisen for discussions. A.C., W.L. and J.L.\ would like to thank the
Aspen
Center for Physics, where part of this work began.
\vskip 4.cm

\appendix A{ Special Geometry}

In this appendix
we first briefly recall the properties of a \K--Hodge manifold
as they are relevant for $N=1$ supergravity. Then we turn to
special \K\ manifolds on whose geometrical structure this
paper is based. We briefly indicate how special geometry
arises from $N=2$ supergravity
and assemble the main formulas
used in the text.
\sect{ \K--Hodge manifolds}
The metric of
an $n$--dimensional \K\ manifold $\cM$ is given by
$$
g_{\a{\bar \b}}(z,\zb)=\del_\a \del_{\bar \b}K(z,\zb)  \, ,
\eqn\metric
$$
where $K(z,\zb)$ is the \K\ potential. Let us introduce the
$1-$form
$\cQ$ defined by
$$
\cQ=- {{\it i}\over 2}\big( \del_\a K  d z^\a-\del_{\bar \a}
 K  d {\bar z}^{\bar \a} \big)  \, .
\eqn\qcon
$$
Under \K\ transformations
$$
K\longrightarrow K+f(z)+\bar f(\zb)\ \,
\eqn\Ktra
$$
$g_{\a\bar\b}$ is left invariant, while $\cal Q$
transforms as a $U
(1)$ connection (\K\ connection):
$$
\cQ \longrightarrow \cQ +{\it d}({\it Im }f ) \, .
\eqn\Qtra
$$
Introducing the \K\ closed $2-$form $\omega$
$$
\omega={\it i} g_{\a{\bar \b}} dz^\a\wedge d{\bar z}^{\bar \b}\, ,
\qquad
 d\omega=0     \, ,
\eqn\omeg
$$
we find
$$
d\cal Q=\omega \, .
\eqn\dQ
$$
Therefore, the first Chern class of the $U(1)$ bundle $L$ whose
connection
is $\cQ$ coincides with the \K\ class $\omega$.
A manifold with this property
is called a \K--Hodge manifold.

A section $\psi(z,\zb)$ of $L$
with  \K\ weight $(p,\bar p)$ is defined
by the
 transformation law
$$
\psi(z,\zb)\longrightarrow \psi(z,\zb)\quad{e^{-\coeff{p}{2}f}}\quad
{e^{-\coeff{\bar p}{2}{\bar f}}}    \, .
\eqn\psit
$$
Accordingly, we define $U(1)$ covariant derivatives
by
$$\eqalign{
D_\a \psi\ &=\ (\del_\a +\coeff{p}{2}\del_\a K)\psi \, ,\cr
D_{\bar \a}\psi\ &=\ (\del_{\bar \a}+\coeff{\bar p}{2}\del_{\bar
\a}K)\psi  \, .
\cr}\eqn\der
$$
A covariantly holomorphic section, satisfying
$
D_{\bar \a}\psi=0
$,
is related to a purely holomorphic field $\tilde \psi$ by
$$
\tilde \psi=e^{\coeff{\bar p}{2}K}\psi    \, .
\eqn\psih
$$
$\tilde \psi$
has weight $(p-\bar p,0)$
and satisfies
$
\del_{\bar \a} \tilde \psi\ =\ 0.
$
The Levi-Civita connections and their curvatures are defined by
$$
\Gamma^\a_{\ \b\c}\ =\ g^{\a{\bar \d}}\del_\b g_{\c{\bar \d}}\, , \qquad
R^\a_{\b{\bar \c}\d}\ =\ \del_{\bar\c} \Gamma^\a_{\b\d} \, .
\eqn\curva
$$
(Analogous formulas hold for the barred quantities $\Gamma^{\bar \a}
_{\ \bar \b \bar \c}$ and $R^{\bar \a}_{\ \bar \b \c \bar \delta}$.)
Thus for a vector $\phi_\a$  of weight $(p,\bar p)$
the  covariant derivatives read
$$\eqalign{
D_\a \phi_\b\ &=\  \big(\del_\a+\coeff{p}{2}\del_\a K\big)\phi_\b-
\Gamma^\c_{\a\b}\phi_\c  \, ,
\cr
D_{\bar \a} \phi_\b\ &=\ \big(\del_{\bar\a}+\coeff{\bar
p}{2}\del_{\bar \a}K\big)\phi_\b \, .\cr}
\eqn\deri
$$

\sect{ Special \K\ manifolds}

The notion of special \K\ geometry first arose in the context of coupling
vector multiplets to $N=2$ supergravity (in four space--time dimensions). It
was shown that the \K\ manifold spanned by the scalar fields $z^\a$ of the
vector multiplets must be suitably restricted as a consequence of $N=2$
supersymmetry \dfnt.
A coordinate--free characterization of such restricted geometry was given in
\fetal\ in the context of $N=2$ supergravity and in \refs{\Strom{,}\CO}\ for a
Calabi-Yau moduli space.

In order to understand how special geometry arises from $N=2$
supersymmetry
let $(\l^{I\a},\l^{\bar \a}_I)$ be the chiral--antichiral components
of the
gaugino field ($I=1,2$ being an $O(2)$ index), and
  $\cA^A_\mu$ ($A=0,1,\dots,n$) the vector superpartners
 and the graviphoton.
On general grounds, their supersymmetry transformation laws  are
$$
\eqalign{
\d_\e \cA_\mu^A &\ =\ f_\a^A \bar \l^{\a I} \gamma_\mu \e^J
\e_{IJ}-2\e_{IJ}
L^A\e^I\psi_\mu^J+{\it h.c.} \, ,\cr
\d_\e \l^{\a I}&\ =\ -2 i g^{\a\bar \b}\e_J\big[ C_{\bar \b \c \d}\bar
\l^{\c I}
\l^{\d J}+C_{\bar \b \bar \c \bar \d} \bar \l^\c_L \l^\d_M
\e^{IL}\e^{JM}\big]   +
\ldots }
\eqn\susy
$$
where the dots stand for terms that are irrelevant for now. Here $\e^I$ and
$\psi^I_\mu$ are the (chiral) supersymmetry parameter and gravitino field
respectively, and $\e_{IJ}$ is the $O(2)$ antisymmetric symbol. The sections
$L^A,\, f^A_\a,\,C_{\bar \a \b\c},\,C_{\bar\a\bar\b\bar\c}$ and their chiral
partners $\bar L^A,\, {\bar f}^A_{\bar \a},\, C_{\a\bar\b\bar\c},\, C_{\a\b\c}$
are a priori unrestricted scalars and tensors whose \K\ weight is fixed by \K\
covariance. The restrictions on the \K\ geometry arise from the on--shell
closure of the above supersymmetry transformation rules. In the superspace
approach, this corresponds to imposing the Bianchi identities on the
supercurvatures. One finds that the closure on $\cA_\mu^A$ implies
$$
\eqalign{
D_{\bar \a} L^A\ =\ 0\, , &\qquad\qquad
 D_\a {\bar L}^A\ =\ 0 \, , \cr
D_\a L^A =\ f_\a^A \, ,&\qquad\qquad
 D_{\bar \a}{\bar L}^A=\bar f_{\bar \a}^A \, , \cr
D_\a \bar f_{\bar \b}^A=\  g_{\a {\bar \b}} \bar L^A\, , &\qquad\qquad D_{\bar
\a}f_\b^A=g_{\a{\bar \b}}{ L}^A        \, .}
\eqn\bia
$$
(Note that the last set of equations is just
 the integrability condition of the second set.)
Closure of the gaugino transformation implies
$$
\eqalign{
C_{\bar\a\b\c}&=C_{\a\bar\b\bar\c}=0 \, ,\cr
D_\a f_\b^A\ &=\ - i C_{\a\b\c}g^{\c{\bar \d}}\bar f_{\bar \d}^A \, ,\cr
D_{\bar \a}\bar f_{\bar \b}^A\ &=\ - i C_{\bar\a \bar\b \bar\c}g^{\bar\c
\d}f_\d^A \, ,\cr
D_{\bar \a} C_{\b\c\d}\ &=\ D_{[\a}C_{\b]\c\d}\ =\ 0 \, ,\cr
D_\a C_{\bar\b \bar\c\bar\d}\ &=\ D_{[\bar\a} C_{\bar\b ]
 \bar\c\bar\d}\
=\ 0 \, ,}
\eqn\cur
$$
as well as
$C_{\a\b\c}$ being a completely symmetric tensor.
As an integrability condition of eq.~\cur, one finds that the curvature
satisfies the following constraint
$$
R_{\bar\a \b \bar\c \d}\ =\
 g_{\bar \a \b}g_{\d \bar \c
}+g_{\bar\a\d}g_{\b\bar\c}-
 C_{\b \d \mu}g^{\mu \bar\mu}
C_{\bar \mu \bar \a \bar \c}   \,.
\eqn\curvb
$$
{}From eq.~\cur\ we also learn
that
$C_{\a\b\c}$ obeys
$$
C_{\a\b\c}=D_\a D_\b D_\c S  \, ,
\eqn\dddS
$$
where $S$ has weight $(2,-2)$.

The above properties lead to the following definition of a special \K\
manifold. A special \K\ manifold is a \K--Hodge manifold for which there exists
a set of $n+1$ sections $L^A(z,\zb)$ and ${\bar L}^A(z,\zb)$ of weight $(1,-1)$
and $(-1,1)$ respectively, satisfying \bia\ and \cur, and a section of weight
$(2,-2)$ ($(-2,2)$) $C_{\a\b\c}$ ($C_{\bar\a\bar\b\bar\c}$) which is completely
symmetric in its indices and satisfies \cur.

Equivalently, a special \K\ manifold can be defined by introducing a 3--index
symmetric tensor $C_{\a\b\c}$ on a \K--Hodge manifold with the properties \cur\
and furthermore restricting the curvature by the constraint \curvb\ . The
existence of the sections $L^A$ and their properties then follow.

The \K\ potential itself is most easily expressed in terms of
holomorphic sections.
By using \psih\
one defines
 $X^A(z)$ and
$W_{\a\b\c}(z)$
 of
\K\ weight $(2,0)$ and  $(4,0)$ respectively:
$$\eqalign{
X^A(z) &= e^{-\coeff{K}{2}}L^A(z,\zb)\ \, ,  \qquad \ \del_{\bar \a}
X^A=0 \, ,\cr
W_{\a\b\c}(z)\ &= e^{-K} C_{\a\b\c}(z,\zb)\, ,\qquad
\del_{\bar\a}W_{\b\c\d}=0 \, . }
\eqn\wsec
$$

We also need to introduce
a functional $F(X^A)$ which is holomorphic and homogeneous of degree
$2$
in the $X^A$:
$$
2F=X^A F_A(X)\, , \qquad F_A\equiv \coeff{\del}{\del X^A} F   \, .
\eqn\hom
$$
In terms of $X^A$ and $F_A$
the \K\ potential which
solves the constraints \cur\ and \curvb\  reads
$$
K(z,\zb) = -\ln i Y \, , \qquad
Y=X^A N_{AB}X^B=X^A\bar F_A-\bar X^A F_A   \,    ,
\eqn\kde
$$
where
$$
N_{AB}=F_{AB}(X)-\bar F_{AB}(\bar X)\, ,\qquad
F_{AB}=\del_A\del_B
F  \, .
\eqn\nij
$$
Furthermore,
$C_{\a\b\c}$ is given by
$$\eqalign{
C_{\a\b\c} &=D_\a D_\b D_\c S=
e^K \del_\a
X^A \del_\b X^B \del_\c X^C F_{ABC}   \, ,
\cr
S &=-{1\over{2}} e^K X^A N_{AB} X^B \, .    }
\eqn\fabc
$$

{}From eqs.~\bia, \cur\ and \wsec-\fabc\
it is straightforward to verify that $X^A$ and
$F_A$ satisfy the same set of constraints.
Therefore we introduce the $(2n+2)$ dimensional row vectors\foot
{We take the expression $(X^A, F_A)$ always as
an abbreviation for $(X^0, X^a, F_a, -F_0)$.}
$$
V\  =\ (X^A, F_A)\ :\equiv\ (X^0,X^\a,F_\a,-F_0)\ ,
\qquad\quad (\a=1,\dots,n)\ .
\eqn\sive
$$
Using \wsec--\fabc\ we  rewrite the
identities \bia\ and \cur\  as follows
$$\eqalign{
D_\a V\ &=\ U_\a \cr
D_\a U_\b\ &=\ - i C_{\a\b\c}g^{\c \bar\d} \bar{U}_{\bar\d} \cr
D_\a {\bar U}_{\bar \b}\ &=\ g_{\a\bar \b} {\bar V} \cr
D_\a \bar V\ &=\ 0  \,  . }
\eqn\sis
$$
It
is this set of constraints  we use in the main text.
Similarly, one derives the constraints including the anti--holomorphic
derivative $D_{\bar \alpha}$.

\ni The \K\ potential can be expressed
in terms of $V$ and $V^\dagger$   as follows:
$$
K = - \ln\left(V(-iQ) V^\dagger \right) \, ,
\eqn\ksymp
$$
which makes its $Sp(2n+2,\IR)$ symmetry manifest. Above, $Q$ is a
symplectic metric which satisfies $Q^2=-1\, ,Q=-Q^T$. Our convention is
$$
 Q\ =\ \pmatrix{
  & & & 1 \cr
  & & -\bfone_n & \cr
  & \bfone_n & & \cr
  -1 & & &\cr}
 \ . \eqn\Qsympdef
$$
Note that the vector $V$ in \sive\ is symplectic with respect to this metric.

An important property which follows from eqs.~\sis\ is that the
connections of special geometry defined in eq.~\deri\
naturally decompose into holomorphic
and non-holomorphic parts \FL. This fact can be displayed by defining
$$
t^a(z)={{X^a}\over{X^0}} \, .
\eqn\speco
$$
In terms of $t^a$ and $X^0$ one finds
$$
\eqalign{
K_\alpha (z,\zb) =\ &
\Kh_\alpha (z) +
\Knh_\alpha (z,\zb)  \, ,  \cr
\Gamma_{\alpha\beta}^\gamma (z,\zb) =\ &
\Gammah_{\alpha\beta}^\gamma (z) +
\T_{\alpha\beta}^\gamma (z,\zb) \, , }
\eqn\conndef
$$
where
$$
\eqalign{
\Knh_\alpha (z,\zb) =&\ e_\alpha^a(z) K_a(z,\zb) \equiv
e_\alpha^a(z)  {\del\over\del t^a} K(t(z),\tb(\zb))          \cr
\Kh_\alpha (z) =& - \del_\alpha \ln X^0 (z) \cr
e_\alpha^a(z) =&\ \del_\alpha t^a (z) \cr
\T_{\alpha\beta}^\gamma (z,\zb) = & \
e_{\alpha}^a e_\beta^b
\del_b g_{a \bar d} g^{-1 \bar d c} e_c^{-1 \gamma}     \cr
\Gammah_{\alpha\beta}^\gamma (z) =&\ (\del_\beta e_\alpha^a)
 e^{-1 \gamma}_a \, .}
\eqn\tconndef
$$
The holomorphic objects  ${\hat K_\a}$ and
 $\hat \Gamma^\a_{\b\c}$ transform as connections
under  \K\ and holomorphic reparametrizations respectively;
moreover $T^\a_{\b\c}$ is a tensor under holomorphic diffeomorphisms
and  $\cK_\a$ is \K\ invariant.
As a consequence one can define
 holomorphic covariant derivatives
in analogy with  \deri\ by
$$
\hat D_\a \phi_\b\ =\ (\del_\a+\coeff{p}{2} \del_\a{\hat K})\phi_\b
-{\hat \Gamma}_{\a\b}^\c \phi_\c \ .
\eqn\holderi
$$
The covariant \pf\ equations precisely use this holomorphic
derivative.

Moreover, $\hat \Gamma$ is a flat connection, i.e.~satisfies
$$
{\hat R}^\c_{\d\a\b}
\equiv \del_\d{\hat\Gamma}^\c_{\a\b}-\del_\a{\hat \Gamma}
^\c_{\d\b}+{\hat \Gamma}^\mu_{\a\b}{\hat
\Gamma}^\c_{\mu\d}-{\hat\Gamma}^\mu_{\d\b}{\hat \Gamma}^\c_{\mu\a}=0\, .
\eqn\hatcurva
$$
The holomorphic metric for which $\hat \Gamma$ is a connection reads
$$
{\hat g}_{\a\b}=e^a_\a e^b_\b \eta_{ab}
\eqn\metric
$$
where $\eta_{ab}$ is a constant (invertible) symmetric matrix. (Note
that $\hat
 g_{\a\b}$ has two holomorphic indices in contrast to the \K\ metric
 $g_{\a\bar\b}$.)

The flat coordinates are exactly the ``special coordinates'' $t^a=z^\a$.
In these coordinates we find
$$
e^a_\a=\delta_\a^a, \qquad {\hat \Gamma}^\d_{\a\b}=0,\qquad {\hat
g}_{\a\b}=\eta_{\a\b}
\eqn\flutstuff
$$
(The gauge choice $X^0=1$ implies $\hat K_\a=0$.)

In terms of $t^a$ one defines the \K\ invariant function
$$
\cF(t^a)\equiv (X^0)^{-2} F(X^A) \, .
\eqn\prepot
$$
The \K\ potential can then be expressed as (up to \K\ freedom)
$$
K\ =\ -{\ln}i\big[ 2 (\cF -\bar \cF)+(\cF_a+\bar \cF_a)(t^a-\bar
t^a)\big] \, .
\eqn\yuk
$$

The special coordinates $t^a$ play the double role of flat coordinates for the
holomorphic geometry with flat connection $\hat \Gamma$ and of ``free falling
frame" coordinates for (non-holomorphic) special geometry. The analogue of
local Lorentz transformations in the free falling frame is given in our case by
the symplectic transformations that relate equivalent patches of special
coordinates.

\appendix B {Remarks on $w_3=0$ and covariantly constant $w_4$.}

We first show  that $w_3=0$ does not imply that the solutions of
eq.~\oddeq\ are equivalent to \Vsol. Let us start from an arbitrary solution
$V= (v_1,v_2,v_3,v_4)$. It is always possible to rescale the entire vector $V$
by $1/v_1$. This leads to $V \rightarrow \tilde V = (1, v_2/v_1, v_3/v_1,
v_4/v_1)$ where $\tilde V$ satisfies an equation \oddeq\ with $a_0=0$. In the
next step we perform the coordinate transformation $z \rightarrow \tilde z =
v_2/v_1$. In these coordinates eq.~\oddeq\ turns into
\def\delt{\tilde{\del}}
\def\at{\tilde{a}}
\def\zt{\tilde{z}}
$$
\left(\delt^4 + \at_3 \delt^3 + \at_2 \delt^2 \right)\tilde V = 0 \, ,
\qquad
\tilde V =(1,\tilde z,f_1(\zt),f_2(\zt))
\eqn\gensol
$$
(Again, we have scaled out $a_4$). The two steps so far can be done for any
fourth--order equation. It is equivalent to fixing the scale (\K)--freedom and
the coordinate frame. In these new coordinates $w_3$ is given by (we drop the
tilde)
$$
w_3 = -\del a_2 - \half a_2 a_3 + \half \del^2 a_3
+ \coeff34 a_3 \del a_3 + \coeff18 a_3^3    \, .
\eqn\wdrei
$$
By writing $a_3 = - 2 \del\ln W$
 and $a_2= b_2 W$, $w_3$ simplifies to
$$
w_3 = W ( \del b_2 - \del^3 W^{-1})\ .
\eqn\wdreip
$$
Thus $w_3=0$ implies the relation
$$
b_2 = \del^2 W^{-1} + c_1                            \, ,
\eqn\wnull
$$
where $c_1$ is a constant.
Inserting \wnull\ into  \gensol\ we find
$$
\left(\del^2 W^{-1} \del^2  + c_1 \del^2 \right) \tilde V = 0  \, .
\eqn\trouble
$$
For $c_1=0$ this is solved by
$$
\del^2 f_1 = W, \qquad \del^2 f_0 = z W
\eqn\czero
$$
which implies
$$
f_0 = z f_1 - 2 f_1 + c_2 z + c_3 \, .
\eqn\solsg
$$
This is precisely what is true in special geometry. However, one can easily
check (by series expansion)
that this does not apply any more if $c \neq 0$. This means that $w_3=0$
does not fully characterize the differential equation \sgeq\ of special
geometry.

Finally, we briefly discuss solutions of the generic fourth--order equation
\wdeq\ with covariantly constant $w_4$:
$$
\hat D\,w_4=0 \ .
\eqn\doubleu
$$
This can easily be solved in special coordinates, where \doubleu\ reduces to
$\del  w_4(t)=0$, by setting
$$
W(t)=e^{\sqrt{5} \a t}\ .
\eqn\yuka
$$
{}From this we obtain
$w_2=-\coeff{5}{2}\a^2$ and $w_4=\a^4$,
and \wdeq\ is solved by first changing to the coordinate system $u(t)$ where
$w_2=0$, that is, where
$$
\{u,t\}=-\coeff{1}{2}\a^2\ .
\eqn\schwar
$$
Then one solves the associated second--order linear differential equation
$$
\theta''-\a^2\theta =0  \Rightarrow \theta_1=e^{\a t}\qquad ,\qquad
 \theta_2=e^{-\a t}\qquad u(t)\equiv {\theta_1
\over{\theta_2}}=e^{2\a t}\ .
\eqn\seco
$$
In the coordinates $u$ we have
$$
w_4(u)=\big({dt\over{du}}\big)^4w_4(t)=\big(\a {dt \over{du}}\big)^4={1\over{16
 u^4}}\ ,
\eqn\coord
$$
and the fourth--order differential equation \wdeq\ becomes:
$$
{\tilde V}^{iv}+{1\over{16 u^4}} {\tilde V}=0 \qquad\qquad {\tilde
 V}=\big(u')\big)^{3/2}V
\eqn\vitilde
$$
It has solutions ${\tilde V}=u^{\b_i}$, where $\b_i$ are the roots of
$$
\b(\b-1)(\b-2)(\b-3)+{1\over 16}=0\ .
\eqn\roots
$$
Altogether we find:
$$
V(t)= e^{3\a t} \big(e^{2\a\b_1 t} , e^{2\a\b_2 t} , e^{2\a\b_3 t} , e^{4\a\b_4
 t} \big)\ .
\eqn\fin
$$
This is similar to the instanton--corrected solution of \CandB, and
more specifically suggests that a covariantly constant $w_4$
characterizes single instantons, in accordance with our considerations
in sect.\ 2.1.

\appendix C{Differential equations for cubic $F$-functions}

It is helpful to first reconsider the first--order system for one variable. We
have seen in sect.~2.2 that for constant Yukawa coupling and in special
coordinates, where $W=1$ and $F=\coeff16t^3$, the matrix connection is given by
the step generator
$$\IGa_w\ =\ \IC\ \equiv\ \pmatrix{
0  & 1 &  0 & 0 \cr
0 & 0  & 1 & 0 \cr
0 & 0 & 0 & 1\cr
0 & 0 & 0 & 0\cr}\ =\ J_-\ ,
\eqn\gnull
$$
of the principal $SL(2)$ subgroup $\cS\subset Sp(4)$.
The diagonal generator $J_0$ belongs to the gauge group.
Thus, in accordance with our considerations in sect 3.1, the moduli space is
$SL(2)/U(1)$.

More generally, all special geometries with a cubic $F$--function
$$
F={1\over3!}\,W_{abc} {X^aX^bX^c\over X^0}\ 
$$
correspond to special, homogeneous K\"ahlerian manifolds, $G/H$
\doubref\dfnt\cubicF. They typically describe moduli spaces of orbifolds.

Let us fix the gauge $X^a\equiv t^a, X^0\equiv1$. Then the flat coordinates
$t^a$ are associated to $G/H$. More precisely, they are associated with the
(mutually commuting) broken raising generators of $G$ in the Cartan-Weyl basis,
and thus they are coordinates of $G^c/B$ (which is, essentially, isomorphic to
$G/H$). Furthermore, the subgroups $H$ act linearly on the coordinates. One may
view these groups as being gauged by the connections
$\Gammah$ in \gammahat. Thus the generalization to many variables is the system
of coupled matrix differential equations
$$
\Big[\,{\bfone}\del_a\ -\ \IC_a\,\Big]\bv\ =\ 0, \eqn\multmatdeq
$$
where $\IC_a$ are the generators of $G/H$
appropriately embedded into $sp(2n+2)$ (with $n=dim_cG/H$).
These equations are solved by each column of the symplectic matrix
$$
\bv\ =\ \exx{t^a\IC_a}\ =\
\pmatrix{ 1 & {t^a} & F_a & F \cr 0 & 1 & F_{ab} &
{t^a}\,F_{ab}- F_b \cr 0 & 0 & 1 & {t^b} \cr 0 & 0 & 0 & 1 \cr  }\ ,
\eqn\Esolu
$$
whose first row gives the ``period'' vector\foot{Note that the components of
$V$ are like elements of some local ring $\Rt$, the structure constants of
which are given by the coset generators $\IC_\a$.}, $V=(1,t^a,\shalf
W_{abc}t^bt^c, \coeff16 W_{abc}t^at^bt^c)$. Observe that $\bv\in G^c/B\cong
G/H$ reflecting the fact that the moduli space is given by $G/H$. Furthermore,
the Yukawa couplings are just the top-bottom components of the triple product
of coset generators (cf. \wabcmat):
$$
W_{abc}\ =\  {\big(\IC_a\IC_b\IC_c\big)_1}^{(2n+2)}
$$

The symplectic embeddings\foot {Such symplectic embeddings have also been
considered in \Fre.} of $G$ generalize the principal embedding of $\cS$. Note
that in order for $F$ to be cubic, the representation of $V$ must be
irreducible with respect to $G$:
$$
R=\underline{2n+2} {\rm\ of\ } Sp(2n+2,\IR) \ \longrightarrow\
r=\underline{2n+2} {\rm\ of\ } G\ ,
$$
so that the top and bottom rows of $\bv$ are highest and lowest weights of $r$.
Otherwise, the action of $\IC_a$ vanishes on some intermediate components of
$\bv$ (the highest weights), which implies that $W_{abc}\equiv0$. For example,
for ${SU(n,1)\over U(n)}$ with $n>1$,
$r=\underline{n+1}\oplus\bar{\underline{n+1}}$ is reducible and accordingly,
$F$ is not cubic but only quadratic.

\ni As an example, consider \Fre
$$
G/H={ SU(3,3) \over {SU(3)\times SU(3)\times U(1)} }\ ,\qquad\qquad
dim_cG/H=n=9\ .\eqn\coset
$$
Here, $G=SU(3,3)$ is maximally embedded in $Sp(20)$ according
to $20=(6\times 6\times 6)_{antisym}$. The  variables $t^a$ correspond to the
broken generators $\IC_a$, which transform as $(3,\bar 3)$ under $SU(3)\otimes
SU(3)$. These matrices are the following nine
commuting generators of $G=SU(3,3)$ in the $20-$dimensional,
threefold totally symmetric representation:
$$
\pmatrix{0 & \delta _i^j \delta_{\bar i} ^{\bar j} & 0 & 0\cr0 & 0 & \e_{ijk}
\e_{\bar i \bar j \bar k} & 0\cr 0 & 0 & 0 &
\delta _i ^j \delta _{\bar i} ^{\bar j} \cr 0& 0 & 0 & 0 }
\eqn\matri
$$
with $i,\bar i,=1,\cdots 3 $.
The local symmetry group $H=U(3)\times U(3)$ is embedded in
$U(9)\times U(1)$.

\refout
\end